\documentclass[longauth, letter]{aa} % for the letters 

\usepackage{graphicx}
\usepackage[breaklinks,pdfnewwindow=true,colorlinks=true,citecolor=blue]{hyperref}
\usepackage{txfonts}
\usepackage{rotating}
\usepackage{amsmath}
\usepackage{longtable}
\usepackage{tablefootnote}
\usepackage{subcaption}
\usepackage{float}
\usepackage{amssymb}
\usepackage{xcolor}
\usepackage{booktabs}
\usepackage[version=3]{mhchem}
\usepackage{mathrsfs}
\usepackage{multirow,bigdelim,dcolumn,booktabs}
\usepackage{graphicx,caption,subcaption}
\usepackage{orcidlink}
\DeclareUnicodeCharacter{03B2}{$\beta$}

\newcommand{\orcid}[1]{\unskip\protect\href{https://orcid.org/#1}{\protect\includegraphics[width=8pt,clip]{logo_orcid}}}
\newif\ifAMStwofonts

\newcommand{\icoms} {iCOMs}

\newcommand{\VirsB}  {$V^{\rm IRS7B}_{\rm sys}$}

\newcommand{\Eup}  {$E_{\rm up}$}
\newcommand{\RCrA}  {CrA~}
\begin{document} 

   \title{FAUST\\ XIII. Dusty cavity and molecular shock driven by IRS7B in the Corona Australis cluster}
   
   \author{G.~Sabatini\inst{1}\orcidlink{0000-0002-6428-9806}
   \and
   {L.~Podio}\inst{1}\orcidlink{0000-0003-2733-5372}\and
   {C.~Codella}\inst{1,2}\orcidlink{0000-0003-1514-3074}\and
   {Y.~Watanabe}\inst{3}\orcidlink{0000-0002-9668-3592}\and
   {M.~De Simone}\inst{1,4}\orcidlink{0000-0001-5659-0140}\and
   {E.~Bianchi}\inst{5}\orcidlink{0000-0001-9249-7082}\and
   {C.~Ceccarelli}\inst{2}\orcidlink{0000-0001-9664-6292}\and
   {C.~J.~Chandler}\inst{6}\orcidlink{0000-0002-7570-5596}\and
   {N.~Sakai}\inst{7}\orcidlink{0000-0002-3297-4497}\and
   {B.~Svoboda}\inst{6}\orcidlink{0000-0002-8502-6431}\and
   {L.~Testi}\inst{1,8}\orcidlink{0000-0003-1859-3070}\and
   {Y.~Aikawa}\inst{9}\orcidlink{0000-0003-3283-6884}\and
   {N.~Balucani}\inst{10}\orcidlink{0000-0001-5121-5683}\and
   {M.~Bouvier}\inst{11}\orcidlink{0000-0003-0167-0746}\and
   {P.~Caselli}\inst{12}\orcidlink{0000-0003-1481-7911}\and
   {E.~Caux}\inst{13}\orcidlink{0000-0002-4463-6663}\and
   {L.~Chahine}\inst{2}\orcidlink{0000-0003-3364-5094}\and
   {S.~Charnley}\inst{14}\orcidlink{0000-0001-6752-5109}\and
   {N.~Cuello}\inst{2}\orcidlink{0000-0003-3713-8073}\and
   {F.~Dulieu}\inst{15}\orcidlink{0000-0001-6981-0421}\and
   {L.~Evans}\inst{16}\orcidlink{0009-0006-1929-3896}\and
   {D.~Fedele}\inst{1}\orcidlink{0000-0001-6156-0034}\and
   {S.~Feng}\inst{17}\orcidlink{0000-0002-4707-8409}\and
   {F.~Fontani}\inst{1,12}\orcidlink{0000-0003-0348-3418}\and
   {T.~Hama}\inst{18,19}\orcidlink{0000-0002-4991-4044}\and
   {T.~Hanawa}\inst{20}\orcidlink{0000-0002-7538-581X}\and
   {E.~Herbst}\inst{21}\orcidlink{0000-0002-4649-2536}\and
   {T.~Hirota}\inst{22}\orcidlink{0000-0003-1659-095X}\and
   {A.~Isella}\inst{23}\orcidlink{0000-0001-8061-2207}\and
   {I.~J\'{i}menez-Serra}\inst{24}\orcidlink{0000-0003-4493-8714}\and
   {D.~Johnstone}\inst{25,26}\orcidlink{0000-0002-6773-459X}\and
   {B.~Lefloch}\inst{27}\orcidlink{0000-0002-9397-3826}\and
   {R.~Le Gal}\inst{2,28}\orcidlink{0000-0003-1837-3772}\and
   {L.~Loinard}\inst{29,30}\orcidlink{0000-0002-5635-3345}\and
   {H.~Baobab Liu}\inst{31}\orcidlink{0000-0003-2300-2626}\and
   {A.~L\'{o}pez-Sepulcre}\inst{2,28}\orcidlink{0000-0002-6729-3640}\and
   {L.~T.~Maud}\inst{4}\orcidlink{0000-0002-7675-3565}\and
   {M.~J.~Maureira}\inst{12}\orcidlink{0000-0002-7026-8163}\and
   {F.~Menard}\inst{2}\orcidlink{0000-0002-1637-7393}\and
   {A.~Miotello}\inst{4}\orcidlink{0000-0002-7997-2528}\and
   {G.~Moellenbrock}\inst{6}\orcidlink{0000-0002-3296-8134}\and
   {H.~Nomura}\inst{32}\orcidlink{0000-0002-7058-7682}\and
   {Y.~Oba}\inst{33}\orcidlink{0000-0002-6852-3604}\and
   {S.~Ohashi}\inst{7}\orcidlink{0000-0002-9661-7958}\and
   {Y.~Okoda}\inst{7,34}\orcidlink{0000-0003-3655-5270}\and
   {Y.~Oya}\inst{33,34}\orcidlink{0000-0002-0197-8751}\and
   {J.~Pineda}\inst{12}\orcidlink{0000-0002-3972-1978}\and
   {A.~Rimola}\inst{36}\orcidlink{0000-0002-9637-4554}\and
   {T.~Sakai}\inst{37}\orcidlink{0000-0003-4521-7492}\and
   {D.~Segura-Cox}\inst{38}\orcidlink{0000-0003-3172-6763}\and
   {Y.~Shirley}\inst{39}\orcidlink{0000-0002-0133-8973}\and
   {C.~Vastel}\inst{13}\orcidlink{0000-0001-8211-6469}\and
   {S.~Viti}\inst{11}\orcidlink{0000-0001-8504-8844}\and
   {N.~Watanabe}\inst{33}\orcidlink{0000-0001-8408-2872}\and
   {Y.~Zhang}\inst{7}\orcidlink{0000-0001-7511-0034}\and
   {Z.~E.~Zhang}\inst{40}\orcidlink{0000-0002-9927-2705}\and
   {S.~Yamamoto}\inst{41}\orcidlink{0000-0002-9865-0970}}
   
      \institute{INAF, Osservatorio Astrofisico di Arcetri, Largo E. Fermi 5, I-50125, Firenze, Italy; \email{giovanni.sabatini@inaf.it} 
      \and Univ. Grenoble Alpes, CNRS, IPAG, 38000 Grenoble, France
      \and Materials Science and Engineering, College of Engineering, Shibaura Institute of Technology, 3-7-5 Toyosu, Koto-ku, Tokyo 135-8548, Japan
      \and European Southern Observatory, Karl-Schwarzschild Str. 2, 85748 Garching bei M¨unchen, Germany
      \and Excellence Cluster ORIGINS, Boltzmannstraße 2D-85748 Garching
      \and National Radio Astronomy Observatory, PO Box O, Socorro, NM 87801, USA
      \and RIKEN Cluster for Pioneering Research, 2-1, Hirosawa, Wako-shi, Saitama 351-0198, Japan
      \and Dipartimento di Fisica e Astronomia “Augusto Righi” Viale Berti Pichat 6/2, Bologna, Italy
      \and Department of Astronomy, The University of Tokyo, 7-3-1 Hongo, Bunkyo-ku, Tokyo 113-0033, Japan
      \and Department of Chemistry, Biology, and Biotechnology, The University of Perugia, Via Elce di Sotto 8, 06123 Perugia, Italy
      \and Leiden Observatory, Leiden University, P.O. Box 9513, 2300 RA Leiden, The Netherlands
      \and Center for Astrochemical Studies, Max-Planck-Institut f¨ur extraterrestrische Physik (MPE), Gieβenbachstr. 1, D-85741 Garching, Germany
      \and IRAP, Univ. de Toulouse, CNRS, CNES, UPS, Toulouse, France
      \and Astrochemistry Laboratory, Code 691, NASA Goddard Space Flight Center, 8800 Greenbelt Road, Greenbelt, MD 20771, USA
      \and CY Cergy Paris Universit´e, Sorbonne Universit´e, Observatoire de Paris, PSL University, CNRS, LERMA, F-95000, Cergy, France
      \and School of Physics and Astronomy, University of Leeds, Leeds LS2 9JT, UK
      \and Department of Astronomy, Xiamen University, Xiamen, Fujian 361005, P. R. China
      \and Komaba Institute for Science, The University of Tokyo, 3-8-1 Komaba, Meguro, Tokyo 153-8902, Japan
      \and Department of Basic Science, The University of Tokyo, 3-8-1 Komaba, Meguro, Tokyo 153-8902, Japan
      \and Center for Frontier Science, Chiba University, 1-33 Yayoi-cho, Inage-ku, Chiba 263-8522, Japan
      \and Department of Chemistry, University of Virginia, McCormick Road, PO Box 400319, Charlottesville, VA 22904, USA
      \and National Astronomical Observatory of Japan, Osawa 2-21-1, Mitaka-shi, Tokyo 181-8588, Japan
      \and Department of Physics and Astronomy, Rice University, 6100 Main Street, MS-108, Houston, TX 77005, USA
      \and Centro de Astrobiolog´ıa (CSIC-INTA), Ctra. de Torrej´on a Ajalvir, km 4, 28850, Torrej´on de Ardoz, Spain
      \and NRC Herzberg Astronomy and Astrophysics, 5071 West Saanich Road, Victoria, BC, V9E 2E7, Canada
      \and Department of Physics and Astronomy, University of Victoria, Victoria, BC, V8P 5C2, Canada
      \and Universit\'{e} de Bordeaux – CNRS Laboratoire d’Astrophysique de Bordeaux, 33600 Pessac, France
      \and Institut de Radioastronomie Millim\'{e}trique, 38406 Saint-Martin d’Heres, France
      \and Instituto de Radioastronom\'{i}a y Astrof\'{i}sica , Universidad Nacional Aut\'{o}noma de M\'{e}xico, A.P. 3-72 (Xangari), 8701, Morelia, Mexico
      \and Instituto de Astronom\'{i}a, Univ. Nacional Autonoma de Mexico, Ciudad Universitaria, A.P. 70-264, Cuidad de Mexico 04510, Mexico
      \and Institute of Astronomy and Astrophysics, Academia Sinica, 11F of Astronomy-Mathematics Building, AS/NTU No.1, Sec. 4, Roosevelt Rd., Taipei 10617, Taiwan, R.O.C.
      \and Division of Science, National Astronomical Observatory of Japan, 2-21-1 Osawa, Mitaka, Tokyo 181-8588, Japan
      \and Institute of Low Temperature Science, Hokkaido University, N19W8, Kita-ku, Sapporo, Hokkaido 060-0819, Japan
      \and Department of Physics, The University of Tokyo, 7-3-1, Hongo, Bunkyo-ku, Tokyo 113-0033, Japan
      \and Yukawa Institute for Theoretical Physics, Kyoto Univ. Oiwake-cho, Kitashirakawa, Sakyo-ku, Kyoto-shi, Kyoto-fu 606-8502, Japan
      \and Departament de Qu´ımica, Universitat Auto`noma de Barcelona, 08193 Bellaterra, Spain
      \and Graduate School of Informatics and Engineering, The University of Electro-Communications, Chofu, Tokyo 182-8585, Japan
      \and Department of Astronomy, The University of Texas at Austin, 2515 Speedway, Austin, Texas 78712, USA
      \and Steward Observatory, 933 N Cherry Ave., Tucson, AZ 85721 USA
      \and Star and Planet Formation Laboratory, RIKEN Cluster for Pioneering Research, Wako, Saitama 351-0198, Japan
      \and SOKENDAI (The Graduate University for Advanced Studies), Shonan Village, Hayama, Kanagawa 240-0193, Japann
      }

   \date{Received~February~14,~2024; Accepted March~23,~2024}

% \abstract{}{}{}{}{} 
% 5 {} token are mandatory
 
  \abstract
   {The origin of the chemical diversity observed around low-mass protostars probably resides in the earliest history of these systems.}
   {We aim to investigate the impact of protostellar feedback on the {chemistry and grain growth in the circumstellar medium of} multiple stellar systems.}
   {In the context of the ALMA Large Program FAUST, we present high-resolution (50 au) observations of CH$_3$OH, H$_2$CO, and SiO and continuum emission at 1.3~mm and 3~mm towards the Corona Australis star cluster.}
   {Methanol emission reveals an arc-like structure at $\sim$1800~au from the protostellar system IRS7B along the direction perpendicular to the major axis of the disc. The arc is located at the edge of two elongated continuum structures that define a cone emerging from IRS7B. The region inside the cone is probed by H$_2$CO, while the eastern wall of the arc shows bright emission in SiO, a typical shock tracer. Taking into account the association with a previously detected radio jet imaged with JVLA at 6 cm, the molecular arc reveals for the first time a bow shock driven by IRS7B and a two-sided dust cavity opened by the mass-loss process. For each cavity wall, we derive an average H$_2$ column density of $\sim$7$\times$10$^{21}$~cm$^{-2}$, a mass of $\sim$9$\times$10$^{-3}$~M$_\odot$, and a lower limit on the dust spectral index of $1.4$.}
   {{These observations provide the first evidence of a shock and a conical dust cavity opened by the jet driven by IRS7B, with important implications for the chemical enrichment and grain growth in the envelope of Solar System analogues}.}


   \maketitle
%
%-------------------------------------------------------------------

\vspace{-15pt}
\section{Introduction}\label{sec1:intro}
\begin{figure*}
   \centering
   \includegraphics[width=0.82\hsize]{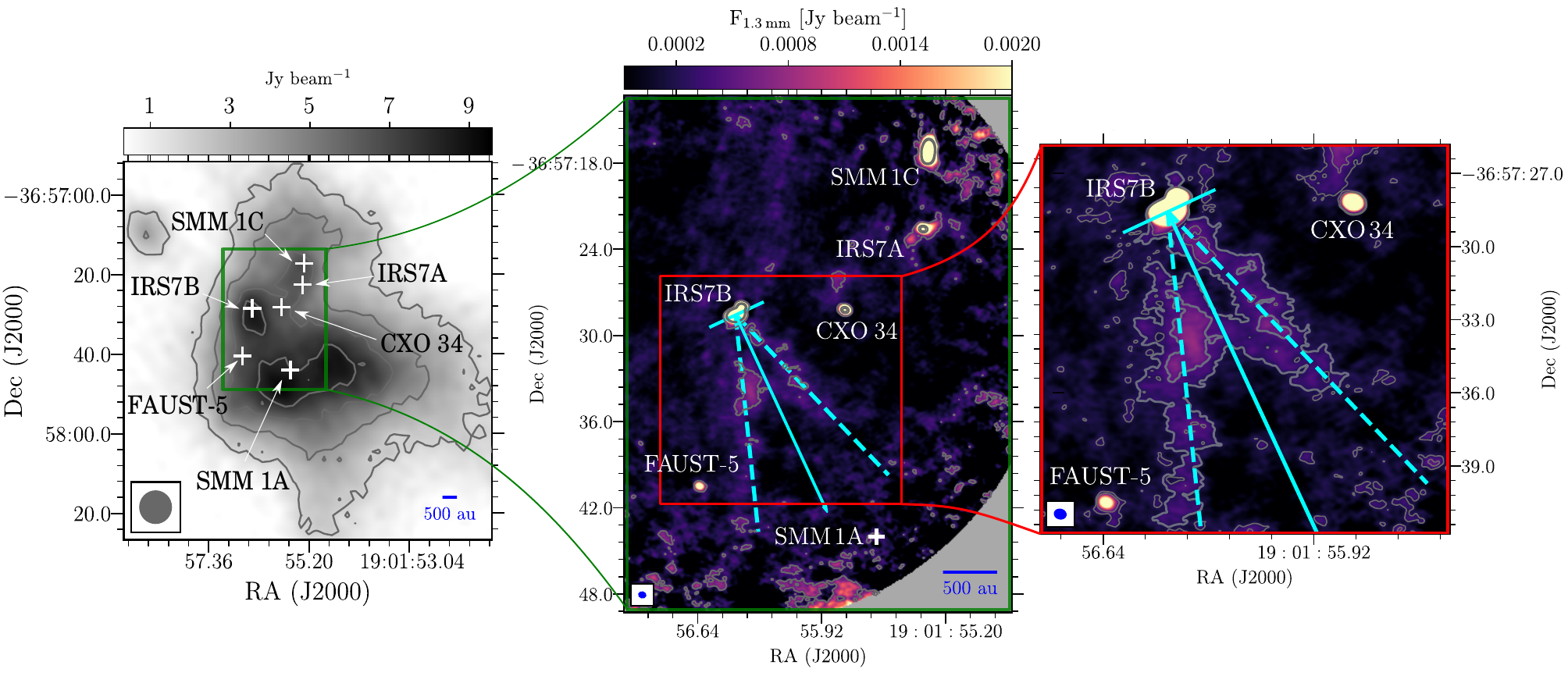}
\caption{The Corona Australis cluster: (Left) SCUBA-2 map at 450~$\mu$m (JCMT archive). Contours are at [12, 18, 30]$\sigma$ where 1$\sigma$~=~0.3 mJy beam$^{-1}$; (Central) ALMA map of continuum emission at 1.3~mm. The first contour and steps are 6$\sigma$ (0.48 mJy beam$^{-1}$) and 70$\sigma$, respectively. (Right) Zoom onto IRS7B, where only the contours at [3,6]$\sigma$ are shown. The detected protostellar objects are labelled. SMM 1A is not detected at 1.3~mm. The cyan line shows the PA of the IRS7B system (i.e. $115^\circ$; see text), while the arrow indicates the perpendicular direction. Dashed lines follow the elongated structures detected at 3$\sigma$. The synthesised beam is shown in the bottom-left corner of each panel.}\label{fig:cont_RCrA}%
\end{figure*}

Protostellar feedback, resulting from the complex interplay between newly forming Sun-like stars and their environment, plays a crucial role in shaping the chemical composition of the interstellar medium (ISM). The injection of energy and momentum leads to a cascade of physical and chemical processes that influence the molecular composition of the ISM \citep[e.g.][as reviews]{Herbst-vanDishoeck09,Frank14, Ceccarelli23}. 
Feedback mechanisms, including stellar winds, jets, {cavities}, and outflows, create a dynamic environment (e.g. shocks) that favours the enrichment of chemical species \citep[e.g.][]{Bachiller2001} and pave the way for the formation of interstellar complex organic molecules (\icoms\footnote{The lower case `i' emphasises that the term `complex' only applies in the context of the ISM.}; i.e. organic molecules with at least six atoms; \citealt{Ceccarelli17}).
Modern interferometers, such as the Atacama Large Millimeter/submillimeter Array (ALMA), enable astronomers to investigate these processes more thoroughly. {However, so far, no clear symmetric dusty cavities opened by protostellar jets have been imaged.}\\
\indent In this context, the ALMA Large Program (LP) Fifty AU STudy of the chemistry in the disk/envelope system of Solar-like protostars (FAUST)\footnote{\url{http://faust-alma.riken.jp}; see \citet{Codella21}.} is designed to characterise the chemical diversity of young solar-like protostars down to a spatial resolution of $\sim$50~au. 
FAUST has shown that the chemical structure of molecular envelopes on scales of $>$1000~au is influenced by protostellar outflow activity \citep[e.g.][]{Okoda21}. New case studies are needed to further investigate the role of protostellar feedback on the different scales of envelope--disc systems.\\
\indent One of the FAUST targets is the nearby Corona Australis (CrA) complex. This region has been the subject of numerous observational campaigns (e.g. \citealt{Knacke73}, \citealt{Harju93} and \citealt{Neuhauser08}). Several Class 0/I protostars have been identified and characterised in terms of both their chemical (e.g. \citealt{Groppi07, Watanabe12} and \citealt{Lindberg14}) and physical properties (e.g. \citealt{Nutter05,Cazzoletti19} and \citealt{Sandell21}). 
IRS7B is the protostar at the centre of the \RCrA cluster, and is classified as a transitional Class 0/Class I object \citep{Groppi07, Watanabe12}. Recently, IRS7B was resolved into a binary system formed by IRS7B-a and IRS7B-b (\citealt{Ohashi23}), both associated with discs aligned with a position angle (PA) of 115$^\circ$ and an inclination of $\sim$ 65$^\circ$ \citep{Takakuwa24}. The cluster also contains CXO~34 and IRS7A, two embedded Class I protostars, and SMM~1C \citep[Class~0;][]{Peterson11, Lindberg15}. These studies also revealed prominent continuum emission from SMM~1A, observed with SCUBA-2 at 450~$\mu$m (Fig.~\ref{fig:cont_RCrA}; left) and located $\sim$2000~au southwest of IRS7B; its nature is still unclear. SMM~1A has been interpreted as either a quiescent prestellar object \citep[e.g.][]{Nutter05} or a filamentary structure affected by external irradiation by the Herbig Ae/Be star R-CrA \citep[e.g.][]{LindbergJorgensen12, Perotti23}.
The present Letter casts a new light on SMM~1A; we present an investigation of the protostellar feedback of IRS7B on the chemical properties of the ISM in the formation of multiple star--disc systems in CrA.

\section{Observations and data reduction}\label{sec2:sample}
The CrA complex was observed as part of the FAUST ALMA LP between October 2018 and September 2019 (2018.1.01205.L; PI: S. Yamamoto; \citealt{Codella21}) at a resolution of $\sim$0$\farcs$4 ($\sim$52~au at the CrA distance of 130 pc; \citealt{Lindberg14}). We observed three spectral setups, one in Band~3 (85-101~GHz) and two in Band~6 (214-234~GHz and 242-262~GHz). Appendix~\ref{App:data} summarises the telescope setups for both the 12m array and the Atacama Compact Array (ACA; Band 6 only). %Different configurations were used, including the 12-m array and the Atacama Compact Array (ACA; Band 6 only). 
The average maximum recoverable scale is $\sim$21$\arcsec$~($\sim$2700 au). All observations were made with a precipitable water vapour of $<$~2.0~mm.\\ %We refer to \cite{Codella21} for a more comprehensive description of the FAUST program.\\
\indent The observations were centred on ($\alpha_{\rm ICRS}, \delta_{\rm ICRS}) = ({\rm 19h01m56s.42}, {\rm -36^\circ57\arcmin28\arcsec.40}$). The absolute flux calibration uncertainty is 10\%. Data were calibrated employing the CASA pipeline~6.2.1-7 (\citealt{CASA_Team_22}), including additional routines to correct for system temperature and spectral normalisation issues\footnote{\url{https://help.almascience.org/index.php?/Knowledgebase/Article/View/419}}, and self-calibrated using line-free channels. The final continuum-subtracted line-cubes were cleaned (Hogbom deconvolver) with \textsc{briggs} weighting scheme with \textsc{robust~$=0.5$} (App.~\ref{App:data}). The typical rms noise is $\sim$0.08~mJy beam$^{-1}$ for continuum and 0.8–3 mJy beam$^{-1}$ per channel for lines. Self-calibration improved the dynamic range of the continuum maps by factors of between 3 and 10 depending on the data set. The primary beam correction was applied to all the
images.

\begin{table*}
        \caption{Properties of the lines detected towards the molecular arc in CrA.}\label{tab:line_properties}
        \setlength{\tabcolsep}{10pt}
        \renewcommand{\arraystretch}{0.7}
        \renewcommand{\tabcolsep}{3pt}
        \centering
        \begin{tabular}{l|cccccc|ccc}
                \toprule
                Transitions & $\nu^{(a)}$ & \Eup$^{(a)}$ & log$_{10}(A_{\rm ul})^{(a)}$ & $g_{\rm u}^{(a)}$ & $\delta V_{\rm chan}^{(a)}$ & $n_{\rm cr}^{(c)}$ & \multicolumn{3}{c}{$\int I$~[K~km s$^{-1}$]$^{(a)}$}\\
        & (MHz) &   (K)    & (s$^{-1}$)  & & (km s$^{-1}$) & (10$^4$~cm$^{-3}$) & ~~~~``A''~~~~ & ~~~~``B''~~~~ & ~~~~``C''~~~~\\
                \midrule
            CH$_3$OH-A~(8$_{0,8}$-7$_{1,7}$) &  95169.391 & 84  & $-5.37$ & 68 & 1.54 & 1.3 -- 1.4 &  8.6 & 4.1    & 6.1\\
            CH$_3$OH-A~(2$_{1,2}$-1$_{1,1}$) &  95914.310 & 22  & $-5.60$ & 20 & 1.54 & 2.0 -- 2.1    &  4.1  & 1.8   & 3.9\\
            CH$_3$OH-A~(3$_{1,3}$-4$_{0,4}$) & 107013.831 & 28  & $-5.51$ & 28 & 0.17 & 3.2 -- 3.5 &  4.5 & $<2.6$ & 2.8\\
            SiO~(5-4)                        & 217104.980 & 31  & $-3.28$ & 11 & 0.51 & 153.1 -- 156.1 &  1.2 & 3.7    & $<0.6$\\
            CH$_3$OH-E~(4$_{2,3}$-3$_{1,2}$) & 218440.063 & 46  & $-4.33$ & 36 & 0.17 & 24.3 -- 23.4   & 19.2 & 8.6    & 9.2\\
            p-H$_2$CO~(3$_{0,3}$-2$_{0,2}$)  & 218222.192 & 21  & $-3.55$ &  7 & 0.17 & 160.1 -- 183.0 & 14.3 & 8.9    & 17.9\\ 
            CH$_3$OH-A~(5$_{1,4}$-4$_{1,3}$) & 243915.788 & 50  & $-4.22$ & 44 & 0.15 & 19.9 -- 21.1    &  --  & 6.9    & 3.9\\

                \bottomrule     
        \end{tabular}
        \tablefoot{$^{(a)}$ Spectroscopic data from The Cologne Database for Molecular Spectroscopy (\citealt{Muller05}); $^{(b)}$ The SiO~(5-4) spectrum is smoothed in velocity by a factor of three to increase sensitivity. $^{(c)}$ Critical densities obtained considering downward collision rates {in a multilevel system} and a temperature range of 50-100 K (Sect.~\ref{sec3.3:RDiagram}). Collisional rates taken from the LAMDA database (\citealt{Schoier05}), based on \cite{Balanca18} for SiO, \cite{Wiesenfeld13} for p-H$_2$CO and \cite{RabliFlower10} for the A- and E-type CH$_3$OH.  $^{(d)}$ We assume an uncertainty of 20\%. Regions `A', `B', and `C' are centred at the peak positions of CH$_3$OH (4$_{2,3}$-3$_{1,2}$), SiO (5-4), and p-H$_2$CO (3$_{0,3}$-2$_{0,2}$), respectively (see Fig.~\ref{fig:moments_zero}).}
\end{table*}

\section{Results}\label{sec3:analisys}
\subsection{Continuum emission}\label{sec3.1:cont}
Figure~\ref{fig:cont_RCrA}~(central) shows the spatial distribution of dust continuum emission at 1.3~mm in the \RCrA field, where four compact sources are detected: IRS7B, CXO~34, IRS7A, and SMM1-C. For these sources, the position of the 1.3~mm continuum peaks derived with a 2D Gaussian fit (Appendix~\ref{app:channels}) matches the coordinates reported by \cite{Lindberg14}. In addition, a fifth source (labelled FAUST-5) is detected for the first time SE of IRS7B. No compact counterpart has been identified in the molecular lines in FAUST, suggesting a potential extragalactic origin.\\
\indent More intriguingly, the continuum emission reveals the existence of two elongated structures  SW of the field. These structures are detected ($>$3$\sigma$) up to $\sim$1000 au (SW) and $\sim$1700 au (S), respectively, from IRS7B. A zoom onto the central region (Fig.~\ref{fig:cont_RCrA}; right) shows that the two structures are symmetric with respect to the normal direction of the disc and extend along the expected direction of the outflow driven by IRS7B, with an opening angle of $\sim$50$^\circ$ (dashed lines). Following \citet{Sabatini23}, we estimate the H$_2$ column density and mass of the cavity {walls, assuming} optically thin dust emission, a dust temperature of 30~K \citep{Perotti23}, {a standard gas-to-dust ratio of 100, and a dust opacity of $\kappa_{\rm 1.3mm}$=~0.9 cm$^2$ g$^{-1}$ \citep{Ossenkopf94}, which are typical values for icy mantle dust grains and well reproduce multi-wavelength observations of low-mass star-forming regions \citep[e.g.][]{Evans01,Shirley05}. The average} $N$(H$_2$) and mass of each cavity wall are $\sim$7$\times$10$^{21}$~cm$^{-2}$ and $\sim$9$\times$10$^{-3}$~M$_\odot$, respectively.
This mass corresponds to $\sim$0.1$\%$ of the IRS7B envelope mass estimated by \cite{vanKempen09}. 
{Taking into account an uncertainty of 10\% in the absolute flux calibration of the ALMA data, $<$15\% in the assumed dust temperature \citep[e.g.][]{Sabatini22} and $\sim$30\% for the dust opacity \citep[e.g.][]{Sanhueza19}, the overall uncertainty for $N$(H$_2$) and the mass of each cavity wall is $\lesssim$35\%.}

\vspace{-10pt}
\subsection{Line emission}\label{sec3.2:lines}
\begin{figure*}
   \centering
   \includegraphics[width=0.75\hsize]{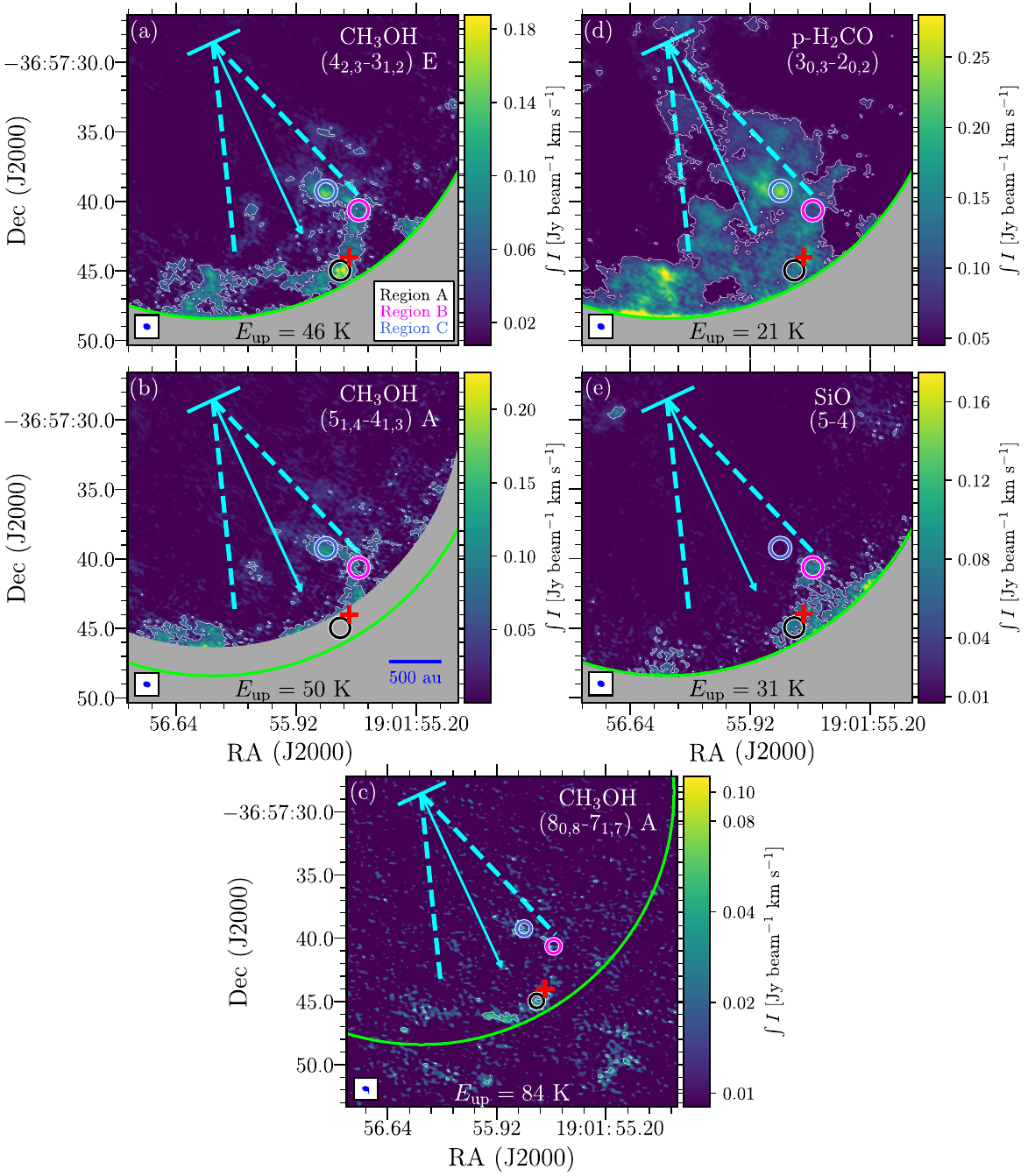}
\caption{Moment 0 of (a) CH$_3$OH-E (4$_{2,3}$-3$_{1,2}$), (b) CH$_3$OH-A~(5$_{1,4}$-4$_{1,3}$), (c) CH$_3$OH-A (8$_{0,8}$-7$_{1,7}$), (d) p-H$_2$CO (3$_{0,3}$-2$_{0,2}$), and (e) SiO (5-4) lines around the molecular arc (integrated from 0 to +12 km~s$^{-1}$). Cyan lines and arrows follow  Fig.~\ref{fig:cont_RCrA}. The white contours mark the 5$\sigma$ emission. Small circles indicate the positions of the brightest spots in CH$_3$OH (4$_{2,3}$-3$_{1,2}$), SiO (5-4), and p-H$_2$CO (3$_{0,3}$-2$_{0,2}$), which we label `A', `B', and `C', respectively. The red cross indicates the position of SMM~1A (Fig.~\ref{fig:cont_RCrA}). The green semicircle shows the ALMA Band~6 FoV, while the grey background delimits the region inside each ALMA pointing.}
\label{fig:moments_zero}%
\end{figure*}

We mapped the molecular emission of five CH$_3$OH lines, associated with a range of {upper level energy}, \Eup, from 21 K to 84 K, para-H$_2$CO~(3$_{0,3}$-2$_{0,2}$), and SiO~(5-4) (see Table~\ref{tab:line_properties}). Figure~\ref{fig:moments_zero} shows the integrated intensity maps (moment 0 maps) of the brightest lines over the velocity range [0, $+12$]~km~s$^{-1}$.\\
\indent Figure~\ref{fig:moments_zero}a shows the spatial distribution of the CH$_3$OH-E~(4$_{2,3}$-3$_{1,2}$), and reveals a molecular-arc structure that is confined within the two elongated structures detected in continuum. The arc has a length of $\sim$3000~au and is located at a projected distance from IRS7B of $\sim$1800~au, close to the coordinates of SMM~1A \citep{Nutter05, ChenArce10}. Additional information is provided by the channel maps (Fig.~\ref{fig:channelmaps}a). The methanol emission extends from +3.5 to +8.0~km~s$^{-1}$ around the systemic velocity of IRS7B (\VirsB~$\sim$~+6~km~s$^{-1}$; \citealt{Ohashi23}), with most of the emission at blueshifted velocities. Similar structures are traced by the other methanol lines in Fig.~\ref{fig:moments_zero}(b, c). The CH$_3$OH-A~(8$_{0,8}$-7$_{1,7}$), detected in Band~3, shows several bright spots consistent with the CH$_3$OH 4$_{2,3}$-3$_{1,2}$ transition, plus fainter spots outside the Band 6 ALMA field of view (FoV; green circle in Fig.~\ref{fig:moments_zero}). This is consistent with the distribution of CH$_3$OH in Band 6 in the ACA-only maps, where the emission is recovered on a larger FoV at the expense of angular resolution (see Appendix~\ref{App:data}). In the CH$_3$OH-A~(5$_{1,4}$-4$_{1,3}$) line, part of the arc structure falls outside the FoV, which is smaller at $\sim$244~GHz compared to that at $\sim$218~GHz. The other methanol lines with low $E_{\rm up}$ (i.e. 22~K and 28~K; Table~\ref{tab:line_properties}) do not allow a clear identification of the arc-like structure. However, these lines are also detected at $>$3$\sigma$ in the spectra integrated on the brightest emission peaks of the arc structure {(Appendix~\ref{app:channels})}.\\
\indent The emission from para-H$_2$CO (3$_{0,3}$-2$_{0,2}$), hereafter referred to as p-H$_2$CO, is more extended than that from methanol; it probes both the arc structure, similarly to CH$_3$OH, and an extended component confined mostly within the boundaries of the cone probed by the dust at 1.3~mm (Fig.~\ref{fig:moments_zero}d). The channel maps show that, at redshifted velocities and close to \VirsB, the p-H$_2$CO emission is extended over a large portion of the FoV, and therefore it is likely contaminated by envelope emission, while it follows the methanol distribution  at blueshifted velocities (Fig.~\ref{fig:channelmaps}b), revealing arc structures.\\
\indent The SiO~(5-4) moment~0~map reveals two emitting regions (Fig.~\ref{fig:moments_zero}e): the first overlaps on the SW side of the molecular arc traced by CH$_3$OH and H$_2$CO; the second is located at the edge of the FoV, beyond the CH$_3$OH molecular arc, at a distance of $>$2100~au from IRS7B. The channel maps show that SiO has the predominant emission component in the blueshifted regime similarly to CH$_3$OH and H$_2$CO  (Fig.~\ref{fig:channel_SOSiO}). This suggests that the blueshifted emission of SiO is also associated with the arc-structure revealed by CH$_3$OH and H$_2$CO (see Sect.~\ref{sec5:conclusions}).\\

\vspace{-20pt}
\subsection{Physical properties of the molecular arc}\label{sec3.3:RDiagram}
\begin{figure}
   \centering
   \includegraphics[width=0.85\hsize]{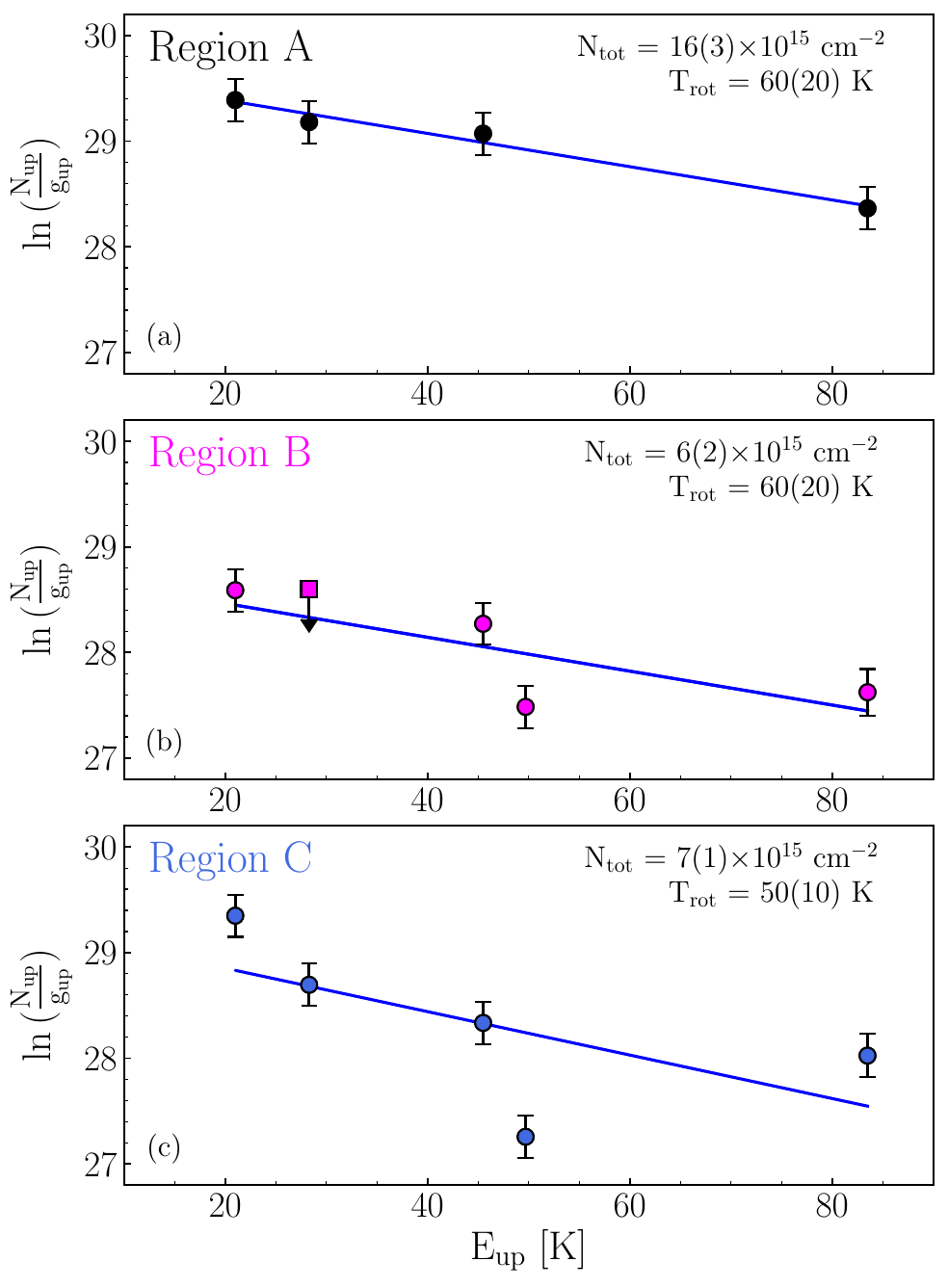}
7\caption{Rotational diagrams of the CH$_3$OH lines (Fig.~\ref{fig:spectra}) for Region: A (in black), B (in magenta), and C (in pale blue). The best-fit values of $T_{\rm rot}$ and total column density are reported.} 
\label{fig:RDiagram}%
\end{figure}

To determine the physical conditions of the emitting gas in the molecular arc, we extracted the spectra of all the observed lines at three positions along the arc structure, labelled `A', `B', and `C', and centred at the peak positions of CH$_3$OH (4$_{2,3}$-3$_{1,2}$), SiO (5-4), and p-H$_2$CO (3$_{0,3}$-2$_{0,2}$), respectively (Fig.~\ref{fig:moments_zero}). Each region has an equivalent area of nine ALMA beams. {The spectra and the line fitting are discussed in Appendix~\ref{app:channels}.}\\
\indent Assuming local thermodynamic equilibrium (LTE) and optically thin emission, we constructed rotational diagrams (RDs) to quantify the column density, $N_{\rm tot}$, and the rotational temperature, $T_{\rm rot}$, of CH$_3$OH  (Fig.~\ref{fig:RDiagram}) observed in the `A', `B', and `C' positions. {It is assumed that the methanol A/E ratio} is equal to unity, which is a good approximation for temperatures of $> 20$~K \citep{Wirstrom11,Carney19}. {We derived the critical densities of the observed lines (Table~\ref{tab:line_properties}) assuming a temperature range of 50-100~K and taking into account downward collision rates in a multi-level system \citep[see][]{Draine11}. The critical densities} range from 10$^4$~cm$^{-3}$ to $\sim$2$\times$10$^6$~cm$^{-3}$ ---below the average H$_2$ number densities determined for SMM~1A (i.e. 10$^6$ cm$^{-3}$; \citealt{LindbergJorgensen12})---, {which supports} the assumption of LTE conditions. The RD analysis is obtained using the line intensity integrated on the blueshifted arc component. Taking into account the calibration uncertainty and the rms of the lines spectra, the errors on the integrated intensities is $\sim$20\%. 
The methanol $N^{\rm CH_3OH}_{\rm tot}$ are between 6$\pm$2$\times$10$^{15}$~cm$^{-2}$ and 16$\pm$3$\times$10$^{15}$~cm$^{-2}$, with $T_{\rm rot}$ of $\sim$50-60~K. In region `A', where methanol peaks, we performed a non-LTE large velocity gradient (LVG) analysis using the \texttt{grelvg} code \citep{Ceccarelli03}. The resulting $T_{\rm rot, LVG}\sim$70-90~K and $N^{\rm CH_3OH}_{\rm tot, LVG}$$\sim$2-6$\times$10$^{15}$~cm$^{-2}$ are in agreement with those derived from the RDs. Indeed the derived line opacities ($\tau<0.5$) and gas densities ($\sim$2$\times$10$^7$~cm$^{-3}$) support the LTE and optically thin RD assumptions. Our results are consistent with the previous estimate by \cite{Perotti23}, who observed SMM~1A, combining Submillimeter Array (SMA) and Atacama Pathfinder Experiment (APEX) observations, and assuming a $T_{\rm rot}\sim$30~K.\\
Assuming the same range of $T_{\rm rot}$ as that obtained for CH$_3$OH, we derive $N_{\rm tot}^{\rm H_2CO}$ $\simeq$ 0.8--2$\times$10$^{14}$~cm$^{-2}$. Our $N_{\rm tot}^{\rm H_2CO}$ values are approximately consistent with the values reported for SMM~1A by \cite{LindbergJorgensen12}, 10$^{13}$--10$^{14}$~cm$^{-2}$, which are based on SMA/APEX observations with a resolution of $\sim$6\arcsec$\times$3\arcsec, and a $T_{\rm rot}$ $\sim$ 48~K. Finally, we derived the total column density of SiO, a typical shock-induced sputtering tracer (see Section~\ref{sec4:discussion_conclusions}), assuming a temperature range of 50-150 K \citep[e.g.][]{Podio21}, obtaining $N_{\rm tot}^{\rm SiO}$ = 1-3 $\times$ 10$^{13}$ cm$^{-2}$ towards the SiO emitting peak (B). 
{The estimated $N_{\rm tot}$, as well as the [CH$_3$OH]/[SiO] and [CH$_3$OH]/[H$_2$CO] abundance ratios ($\sim$250-600, and $\sim$40-130, respectively), are in agreement with the values derived in shocked regions around low-mass protostars (e.g. \citealt{Bachiller1997, Cuadrado17, Podio21, deSimone22}).}

\vspace{-10pt}
\section{Discussion}\label{sec4:discussion_conclusions}
\begin{figure*}
   \centering
   \includegraphics[width=0.8\hsize]{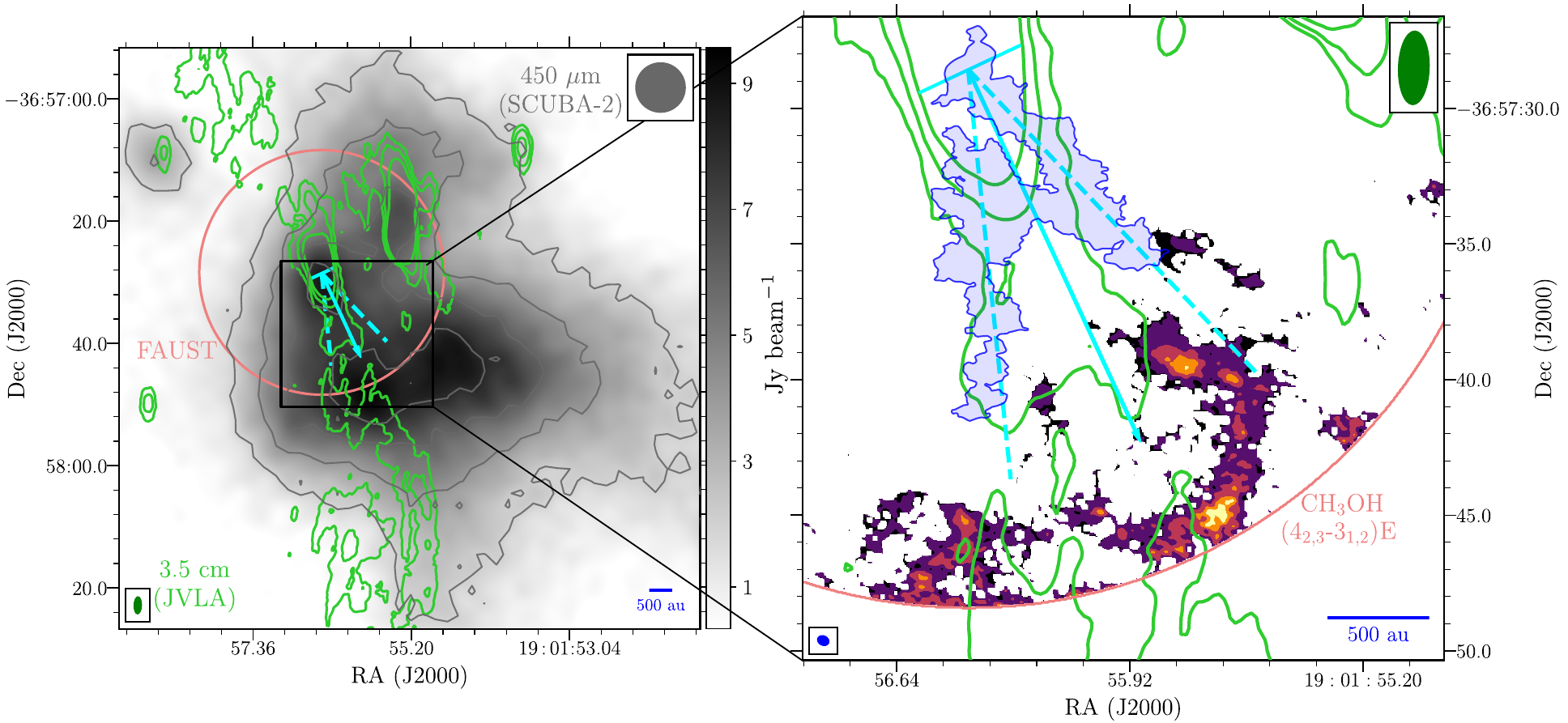}
\caption{Comparison between JVLA, SCUBA, and ALMA-FAUST observations in CrA: (Left) JVLA 3.5~cm map (green contours; \citealt{Liu14}) superimposed on the SCUBA-2 map at 450~$\mu$m (colour map and grey contours; JCMT archive). The green contours are at [5, 10, 20]$\sigma$ where 1$\sigma$~=~8 $\mu$Jy beam$^{-1}$, whilst the grey ones are the same as Fig.~\ref{fig:cont_RCrA}. (Right) Zoom onto the arc (Fig.~\ref{fig:moments_zero}a). The blue-shaded area marks the 3$\sigma$ thermal dust continuum emission taken from Fig.~\ref{fig:cont_RCrA}. The ALMA, JVLA, and SCUBA beams are shown in these panels in blue, green, and grey, respectively.}\label{fig:JVLA}%
\end{figure*}

The ALMA-FAUST data reveal an extended molecular arc-structure traced by CH$_3$OH, where emission due to H$_2$CO and SiO has also been detected. The arc is located in the SW region of the \RCrA cluster and matches the coordinates of SMM~1A. The RD analysis of the CH$_3$OH lines supports that the arc is a coherent structure with relatively constant $N_{\rm tot}^{\rm CH_3OH}$ and $T_{\rm rot}^{\rm CH_3OH}$ . The H$_2$CO distribution shows that the arc is connected to the IRS7B system at the \RCrA cluster centre.

\subsection{{Physical origin of the molecular arc in CrA}}
\indent Methanol is one of the most abundant \icoms~detected in star-forming regions and is considered a key precursor of many prebiotic compounds in space \citep[e.g.][]{Herbst-vanDishoeck09, Ceccarelli23}. The main chemical process for synthesising CH$_3$OH is CO hydrogenation on grains, in which CO is converted to H$_2$CO and then to methanol. This process is extremely efficient in solid phase during the early stages of star formation \citep[e.g.][]{WatanabeKouchi02, Fuchs09, Santos22} when the gas temperatures are $<$20~K, $n(\rm H_2)> 10^{4}$~cm$^{-3}$, and CO is strongly depleted at the surface of dust grains (e.g. \citealt{Caselli99, Bergin02, Sabatini19, Sabatini21, Sabatini22}). The detection of methanol in the arc-structure in CrA can be associated with either ($i$) {external irradiation \citep[see e.g.][]{Cuadrado17}} or ($ii$) shocked regions \citep[e.g.][and references therein]{Ceccarelli23}.\\
\indent The first interpretation was proposed by \cite{LindbergJorgensen12}, who derived a map of H$_2$CO rotational temperature based on SMA/APEX observations with a resolution of 6\arcsec$\times$3$\arcsec$. 
The authors found $T_{\rm rot}^{\rm H_2CO}$ in the 60-100~K range, and interpreted this high value as a consequence of external illumination by the Herbig Ae/Be star R~CrA located at $\sim$50\arcsec~in the NW direction (i.e. $\sim$6500~au at the cluster distance). The interstellar radiation field in SMM~1A would increase by a factor of 750 to 3000 when assuming a luminosity of 100-200~L$_\odot$ for R~CrA, respectively. This picture is in agreement with the recent estimates of \cite{Sissa19}, who report an R~CrA luminosity of $\sim$130~L$_\odot$, and with the relatively small full width at half maximums (FWHMs)\ of the CH$_3$OH and H$_2$CO lines ($<$1.2~km~s$^{-1}$) fitted in Fig.~\ref{fig:spectra}. Nevertheless, the 50 au angular resolution of the FAUST data enabled us to resolve the arc structure with unprecedented accuracy, that is, a factor of $\sim$10 greater accuracy than in \citealt{LindbergJorgensen12}. The newly discovered arc morphology found in SMM~1A would suggest a source of illumination in the southwest (outside the ALMA FoV) or in the cluster centre. Remarkably, no major illumination source is found in either direction, as only the IRS7B system ($L_{\rm bol}\sim 4.6$~L$_\odot$; \citealt{Lindberg14}) and the prestellar source SMM~1As ($L_{\rm bol}\sim 0.04$~L$_\odot$; \citealt{Sicilia-Aguilar13}) have been reported in the literature within 50\arcsec~of SMM~1A.\\
\indent On the other hand, the SiO detection at the edge of the CH$_3$OH arc supports the second scenario, in which the arc is the fingerprint of a shocked region. Indeed, SiO is mainly formed by the shock-induced sputtering and/or shattering of the grain mantles and refractory cores \citep[e.g.][]{Caselli97,Gusdorf08a,Gusdorf08b,Jimenez-Serra08,Guillet11}. In this case, the arc-shape suggests the occurrence of a jet driven by IRS7B towards the SW. Moreover, the molecular arc revealed by FAUST is blueshifted, in agreement with the disc flaring reported by \citet{Takakuwa24}, indicating that the near disc-side is to the NE, and therefore the blueshifted jet is towards the SW.\\
\indent Evidence of outflowing material in \RCrA was first suggested by \cite{Levreault88} based on $^{12}$CO~(2-1) data collected with the Millimeter Wave Observatory (MWO). Although the coarse resolution of the MWO (1\farcs3) prevented the identification of the driving source(s), the blueshifted wing of the outflow is aligned along the direction connecting IRS7B with SMM~1A. 
Radio continuum observations with ATCA\footnote{Australia Telescope Compact Array (ATCA)} at 3, 6, and 20 cm revealed a bipolar structure in the NE-SW direction centred on IRS7B  \citep{Harju01,Choi08,Miettinen08}. More recently, \cite{Liu14} also reported an extended ($\sim$70\arcsec~in the SMM~1A direction) bipolar radio jet detected at 3.5~cm with the Karl G. Jansky Very Large Array (JVLA; beam $\sim$4\arcsec$\times$2$\arcsec$). In Fig.~\ref{fig:JVLA} (left panel), we show the JVLA map (green contrours; \citealt{Liu14}) superimposed on the SCUBA-2 map at 450~$\mu$m (grey scale\footnote{Reduced data can be found at the Canadian Astronomy Data Centre (Project-ID: MJLSG35; \url{https://www.cadc-ccda.hia-iha.nrc-cnrc.gc.ca/en/search/}).}). 
What originally appeared in the SCUBA-2 data as an overdensity of cold gas \citep[e.g.][]{Nutter05, Chen10} overlaps with both ($a$) the direction of propagation of the radio-jet discovered by \cite{Liu14} and ($b$) the molecular arc traced by methanol (see Fig.~\ref{fig:JVLA}, right).
Moreover, the region in which the radio-jet would collide with the arc also corresponds to the zone in which the $T_{\rm rot}^{\rm H_2CO}$ derived in \cite{LindbergJorgensen12} reaches its maximum values. In conclusion, based on the arc-shape, the association with the radio jet detected at centimetre wavelengths, and the detection of a typical shock tracer, namely SiO, we conclude that the observed molecular arc is most likely the signature of a bow-shock driven by IRS7B.

\subsection{{First imaging of symmetric dusty cavity walls}}
In light of these findings, the `V'-shaped structure detected in continuum (Fig.~\ref{fig:JVLA}, right) traces the dust emission from the cavity walls opened by the IRS7B outflow. This is in agreement with the direction of the JVLA radio jet with the orientation of the cavity observed by ALMA (Fig.~\ref{fig:JVLA}). To our knowledge, this is the first {imaging} of symmetric cavity walls of dust opened by a jet (on scales of 50~au) in a low-mass star-forming system.\\
\indent {This provides a unique opportunity to study the dust properties at the edge of an outflow cavity. Under the Rayleigh–Jeans approximation, both the size and the properties of interstellar dust grains can be inferred from the spectral index of the dust spectral energy distribution at (sub)millimetre wavelengths ($\alpha$). 
We measure $\alpha$ in the cavity walls using Band 6 (1.3 mm) and Band 3 (3mm) continuum maps. In Band 3, the cavity walls are not detected, and so we derive a lower limit $\alpha > 1.4$.}\\ %
\indent The spectral index for typical ISM grains is $\alpha_{\rm ISM}$$\sim$3.7 \citep{Testi14}, while low values (down to $\alpha$$\sim$1.5) have been observed in the inner few hundred astronomical units (au) around Class 0/I sources. Low $\alpha$ values can be due to dust growth \citep[e.g.][]{Chiang12, Miotello14, Galametz19}, while recent theoretical calculations and observations show that large grains in the envelope may be the result of transport from the site of growth, the inner dense disc, to the envelope via magnetohydrodynamical (MHD) winds \citep[][]{Wong16, Giacalone19, Tsukamoto21,Cacciapuoti23}. In this context, the estimated $\alpha>1.4$ allows the possibility that high-density cavity walls, like those in CrA, are the sites for in situ dust growth in protostellar envelopes.

\section{Conclusions}\label{sec5:conclusions}
In this Letter, we present high-resolution (50 au) observations of  continuum (1.3~and 3~mm) and molecular (CH$_3$OH, H$_2$CO and SiO) emission towards the CrA cluster in the context of the ALMA FAUST LP. 
Our analysis indicates that SMM~1A, previously identified as an extended continuum structure, is associated with a shock driven by IRS7B, and mapped in CH$_3$OH, H$_2$CO, and SiO, and with a conical dusty cavity opened by the mass-loss process. 
We estimated the H$_2$ column density ($\sim$7$\times$10$^{21}$~cm$^{-2}$) and mass ($\sim$9$\times$10$^{-3}$~M$_\odot$) in the cavity walls, and a lower limit for the dust spectral index ($\alpha > 1.4$), which could imply the presence of millimetre-sized grains.
Based on these results, we conclude that the CrA cluster may be a unique laboratory with which to investigate and test models of dust grain growth in the envelope.\\
\indent Additional higher-sensitivity and higher-resolution observations of shock tracers, such as SiO and CO, are needed to reveal the IRS7B jet. The  discovery of a twin system in IRS7B \citep{Ohashi23} opens the possibility that there are two jets. Therefore, the SiO knot could be due to precession. Further mapping of CH$_3$OH emission in the SW region around SMM~1A and beyond the FAUST FoV is also needed to reveal additional interactions between the jet and the CrA envelope on a larger scale;  our Band-3 maps already suggest their existence. This would allow us to determine the extent  to which the jet influences the chemical composition of the \RCrA cluster. {Ultimately, a more accurate estimate of $\alpha$ will be made possible with more sensitive observations at longer wavelengths, allowing us to constrain the properties and size distribution of the dust grains in the cavity walls and to probe for the formation and/or entrainment of large grains in the envelope.}\\

\begin{acknowledgements}
The authors thank the anonymous Referee for suggestions that improved the manuscript, and A. Garufi for the fruitful discussion and feedback on the properties of interstellar dust. GS, LP and CC acknowledge the project PRIN-MUR 2020 MUR BEYOND-2p (``Astrochemistry beyond the second period elements'', Prot. 2020AFB3FX), the PRIN MUR 2022 FOSSILS (Chemical origins: linking the fossil composition of the Solar System with the chemistry of protoplanetary discs, Prot. 2022JC2Y93), the project ASI-Astrobiologia 2023 MIGLIORA (Modeling Chemical Complexity, F83C23000800005), and the INAF-GO 2023 fundings PROTO-SKA (Exploiting ALMA data to study planet forming disks: preparing the advent of SKA, C13C23000770005). GS acknowledges the INAF-Minigrant 2023 TRIESTE (``TRacing the chemIcal hEritage of our originS: from proTostars to planEts''; PI: G. Sabatini). LP acknowledges the INAF Mini-Grant 2022 “Chemical Origins” (PI: L. Podio). EB acknowledges support from the Deutsche Forschungsgemeinschaft (DFG, German Research Foundation) under German´s Excellence Strategy – EXC 2094 – 390783311. MB and SV acknowledge support from the European Research Council (ERC) Advanced Grant MOPPEX 833460. SBC was supported by the NASA Planetary Science Division Internal Scientist Funding Program through the Fundamental Laboratory Research work package (FLaRe). LL acknowledges the support of DGAPA PAPIIT grants IN108324 and IN112820 and CONACyT-CF grant 263356. IJ-S acknowledges funding from grants No. PID2019-105552RB-C41 and PID2022-136814NB-I00 from the Spanish Ministry of Science and Innovation/State Agency of Research MCIN/AEI/10.13039/501100011033 and by “ERDF A way of making Europe”. This Letter makes use of the following ALMA data: ADS/JAO.ALMA\#2018.1.01205.L (PI: S. Yamamoto). ALMA is a partnership of the ESO (representing its member states), the NSF (USA) and NINS (Japan), together with the NRC (Canada) and the NSC and ASIAA (Taiwan), in cooperation with the Republic of Chile. The Joint ALMA Observatory is operated by the ESO, the AUI/NRAO, and the NAOJ.
\end{acknowledgements}


\bibliographystyle{aa} % style aa.bst
\bibliography{mybib_GAL}

\begin{thebibliography}{73}
\expandafter\ifx\csname natexlab\endcsname\relax\def\natexlab#1{#1}\fi

\bibitem[{{Bachiller} \& {P{\'e}rez Guti{\'e}rrez}(1997)}]{Bachiller1997}
{Bachiller}, R. \& {P{\'e}rez Guti{\'e}rrez}, M. 1997, \apjl, 487, L93

\bibitem[{{Bachiller} {et~al.}(2001){Bachiller}, {P{\'e}rez Guti{\'e}rrez}, {Kumar}, \& {Tafalla}}]{Bachiller2001}
{Bachiller}, R., {P{\'e}rez Guti{\'e}rrez}, M., {Kumar}, M.~S.~N., \& {Tafalla}, M. 2001, \aap, 372, 899

\bibitem[{{Balan{\c{c}}a} {et~al.}(2018){Balan{\c{c}}a}, {Dayou}, {Faure}, {Wiesenfeld}, \& {Feautrier}}]{Balanca18}
{Balan{\c{c}}a}, C., {Dayou}, F., {Faure}, A., {Wiesenfeld}, L., \& {Feautrier}, N. 2018, \mnras, 479, 2692

\bibitem[{{Bergin} {et~al.}(2002){Bergin}, {Alves}, {Huard}, \& {Lada}}]{Bergin02}
{Bergin}, E.~A., {Alves}, J., {Huard}, T., \& {Lada}, C.~J. 2002, \apjl, 570, L101

\bibitem[{{Cacciapuoti} {et~al.}(2024){Cacciapuoti}, {Testi}, {Podio}, {Codella}, {Maury}, {De Simone}, {Hennebelle}, {Lebreuilly}, {Klessen}, \& {Molinari}}]{Cacciapuoti23}
{Cacciapuoti}, L., {Testi}, L., {Podio}, L., {et~al.} 2024, \apj, 961, 90

\bibitem[{{Carney} {et~al.}(2019){Carney}, {Hogerheijde}, {Guzm{\'a}n}, {Walsh}, {{\"O}berg}, {Fayolle}, {Cleeves}, {Carpenter}, \& {Qi}}]{Carney19}
{Carney}, M.~T., {Hogerheijde}, M.~R., {Guzm{\'a}n}, V.~V., {et~al.} 2019, \aap, 623, A124

\bibitem[{{Caselli} {et~al.}(1997){Caselli}, {Hartquist}, \& {Havnes}}]{Caselli97}
{Caselli}, P., {Hartquist}, T.~W., \& {Havnes}, O. 1997, \aap, 322, 296

\bibitem[{{Caselli} {et~al.}(1999){Caselli}, {Walmsley}, {Tafalla}, {Dore}, \& {Myers}}]{Caselli99}
{Caselli}, P., {Walmsley}, C.~M., {Tafalla}, M., {Dore}, L., \& {Myers}, P.~C. 1999, \apjl, 523, L165

\bibitem[{{Cazzoletti} {et~al.}(2019){Cazzoletti}, {Manara, C. F.}, {Baobab Liu, H.}, {van Dishoeck, E. F.}, {Facchini, S.}, {Alcalà, J. M.}, {Ansdell, M.}, {Testi, L.}, {Williams, J. P.}, {Carrasco-González, C.}, {Dong, R.}, {Forbrich, J.}, {Fukagawa, M.}, {Galván-Madrid, R.}, {Hirano, N.}, {Hogerheijde, M.}, {Hasegawa, Y.}, {Muto, T.}, {Pinilla, P.}, {Takami, M.}, {Tamura, M.}, {Tazzari, M.}, \& {Wisniewski, J. P.}}]{Cazzoletti19}
{Cazzoletti}, P., {Manara, C. F.}, {Baobab Liu, H.}, {et~al.} 2019, \aap, 626, A11

\bibitem[{{Ceccarelli} {et~al.}(2017){Ceccarelli}, {Caselli}, {Fontani}, {Neri}, {L{\'o}pez-Sepulcre}, {Codella}, {Feng}, {Jim{\'e}nez-Serra}, {Lefloch}, {Pineda}, {Vastel}, {Alves}, {Bachiller}, {Balucani}, {Bianchi}, {Bizzocchi}, {Bottinelli}, {Caux}, {Chac{\'o}n-Tanarro}, {Choudhury}, {Coutens}, {Dulieu}, {Favre}, {Hily-Blant}, {Holdship}, {Kahane}, {Jaber Al-Edhari}, {Laas}, {Ospina}, {Oya}, {Podio}, {Pon}, {Punanova}, {Quenard}, {Rimola}, {Sakai}, {Sims}, {Spezzano}, {Taquet}, {Testi}, {Theul{\'e}}, {Ugliengo}, {Vasyunin}, {Viti}, {Wiesenfeld}, \& {Yamamoto}}]{Ceccarelli17}
{Ceccarelli}, C., {Caselli}, P., {Fontani}, F., {et~al.} 2017, \apj, 850, 176

\bibitem[{{Ceccarelli} {et~al.}(2023){Ceccarelli}, {Codella}, {Balucani}, {Bockelee-Morvan}, {Herbst}, {Vastel}, {Caselli}, {Favre}, {Lefloch}, {Oberg}, \& {Yamamoto}}]{Ceccarelli23}
{Ceccarelli}, C., {Codella}, C., {Balucani}, N., {et~al.} 2023, in ASP Conference Series, Vol. 534, ASP Conference Series, ed. S.~{Inutsuka}, Y.~{Aikawa}, T.~{Muto}, K.~{Tomida}, \& M.~{Tamura}, 379

\bibitem[{{Ceccarelli} {et~al.}(2003){Ceccarelli}, {Maret}, {Tielens}, {Castets}, \& {Caux}}]{Ceccarelli03}
{Ceccarelli}, C., {Maret}, S., {Tielens}, A.~G.~G.~M., {Castets}, A., \& {Caux}, E. 2003, \aap, 410, 587

\bibitem[{Chen {et~al.}(2010)Chen, Liu, Su, \& Zhang}]{Chen10}
Chen, H.-R., Liu, S.-Y., Su, Y.-N., \& Zhang, Q. 2010, ApJ, 713, L50

\bibitem[{{Chen} \& {Arce}(2010)}]{ChenArce10}
{Chen}, X. \& {Arce}, H.~G. 2010, \apjl, 720, L169

\bibitem[{{Chiang} {et~al.}(2012){Chiang}, {Looney}, \& {Tobin}}]{Chiang12}
{Chiang}, H.-F., {Looney}, L.~W., \& {Tobin}, J.~J. 2012, \apj, 756, 168

\bibitem[{{Choi} {et~al.}(2008){Choi}, {Hamaguchi}, {Lee}, \& {Tatematsu}}]{Choi08}
{Choi}, M., {Hamaguchi}, K., {Lee}, J.-E., \& {Tatematsu}, K. 2008, \apj, 687, 406

\bibitem[{{Codella} {et~al.}(2021){Codella}, {Ceccarelli}, {Chandler}, {Sakai}, {Yamamoto}, \& {FAUST Team}}]{Codella21}
{Codella}, C., {Ceccarelli}, C., {Chandler}, C., {et~al.} 2021, FSPAS, 8, 227

\bibitem[{{Cuadrado} {et~al.}(2017){Cuadrado}, {Goicoechea}, {Cernicharo}, {Fuente}, {Pety}, \& {Tercero}}]{Cuadrado17}
{Cuadrado}, S., {Goicoechea}, J.~R., {Cernicharo}, J., {et~al.} 2017, \aap, 603, A124

\bibitem[{{De Simone} {et~al.}(2022){De Simone}, {Codella}, {Ceccarelli}, {L{\'o}pez-Sepulcre}, {Neri}, {Rivera-Ortiz}, {Busquet}, {Caselli}, {Bianchi}, {Fontani}, {Lefloch}, {Oya}, \& {Pineda}}]{deSimone22}
{De Simone}, M., {Codella}, C., {Ceccarelli}, C., {et~al.} 2022, \mnras, 512, 5214

\bibitem[{{Draine}(2011)}]{Draine11}
{Draine}, B.~T. 2011, {Physics of the Interstellar and Intergalactic Medium} by Bruce T. Draine. Princeton University Press

\bibitem[{{Evans} {et~al.}(2001){Evans}, {Rawlings}, {Shirley}, \& {Mundy}}]{Evans01}
{Evans}, Neal~J., I., {Rawlings}, J. M.~C., {Shirley}, Y.~L., \& {Mundy}, L.~G. 2001, \apj, 557, 193

\bibitem[{{Frank} {et~al.}(2014){Frank}, {Ray}, {Cabrit}, {Hartigan}, {Arce}, {Bacciotti}, {Bally}, {Benisty}, {Eisl{\"o}ffel}, {G{\"u}del}, {Lebedev}, {Nisini}, \& {Raga}}]{Frank14}
{Frank}, A., {Ray}, T.~P., {Cabrit}, S., {et~al.} 2014, in Protostars and Planets VI, ed. H.~{Beuther}, R.~S. {Klessen}, C.~P. {Dullemond}, \& T.~{Henning}, 451--474

\bibitem[{{Fuchs} {et~al.}(2009){Fuchs}, {Cuppen}, {Ioppolo}, {Romanzin}, {Bisschop}, {Andersson}, {van Dishoeck}, \& {Linnartz}}]{Fuchs09}
{Fuchs}, G.~W., {Cuppen}, H.~M., {Ioppolo}, S., {et~al.} 2009, \aap, 505, 629

\bibitem[{{Galametz} {et~al.}(2019){Galametz}, {Maury}, {Valdivia}, {Testi}, {Belloche}, \& {Andr{\'e}}}]{Galametz19}
{Galametz}, M., {Maury}, A.~J., {Valdivia}, V., {et~al.} 2019, \aap, 632, A5

\bibitem[{{Giacalone} {et~al.}(2019){Giacalone}, {Teitler}, {K{\"o}nigl}, {Krijt}, \& {Ciesla}}]{Giacalone19}
{Giacalone}, S., {Teitler}, S., {K{\"o}nigl}, A., {Krijt}, S., \& {Ciesla}, F.~J. 2019, \apj, 882, 33

\bibitem[{{Ginsburg} {et~al.}(2022){Ginsburg}, {Sokolov}, {de Val-Borro}, {Rosolowsky}, {Pineda}, {Sip{\H{o}}cz}, \& {Henshaw}}]{Ginsburg22}
{Ginsburg}, A., {Sokolov}, V., {de Val-Borro}, M., {et~al.} 2022, \aj, 163, 291

\bibitem[{{Groppi} {et~al.}(2007){Groppi}, {Hunter}, {Blundell}, \& {Sandell}}]{Groppi07}
{Groppi}, C.~E., {Hunter}, T.~R., {Blundell}, R., \& {Sandell}, G. 2007, \apj, 670, 489

\bibitem[{{Guillet} {et~al.}(2011){Guillet}, {Pineau Des For{\^e}ts}, \& {Jones}}]{Guillet11}
{Guillet}, V., {Pineau Des For{\^e}ts}, G., \& {Jones}, A.~P. 2011, \aap, 527, A123

\bibitem[{{Gusdorf} {et~al.}(2008{\natexlab{a}}){Gusdorf}, {Cabrit}, {Flower}, \& {Pineau Des For{\^e}ts}}]{Gusdorf08a}
{Gusdorf}, A., {Cabrit}, S., {Flower}, D.~R., \& {Pineau Des For{\^e}ts}, G. 2008{\natexlab{a}}, \aap, 482, 809

\bibitem[{{Gusdorf} {et~al.}(2008{\natexlab{b}}){Gusdorf}, {Pineau Des For{\^e}ts}, {Cabrit}, \& {Flower}}]{Gusdorf08b}
{Gusdorf}, A., {Pineau Des For{\^e}ts}, G., {Cabrit}, S., \& {Flower}, D.~R. 2008{\natexlab{b}}, \aap, 490, 695

\bibitem[{{Harju} {et~al.}(1993){Harju}, {Haikala}, {Mattila}, {Mauersberger}, {Booth}, \& {Nordh}}]{Harju93}
{Harju}, J., {Haikala}, L.~K., {Mattila}, K., {et~al.} 1993, \aap, 278, 569

\bibitem[{{Harju} {et~al.}(2001){Harju}, {Higdon}, {Lehtinen}, \& {Juvela}}]{Harju01}
{Harju}, J., {Higdon}, J.~L., {Lehtinen}, K., \& {Juvela}, M. 2001, in ASP Conference Series, Vol. 235, Science with the Atacama Large Millimeter Array, ed. A.~{Wootten}, 125

\bibitem[{{Herbst} \& {van Dishoeck}(2009)}]{Herbst-vanDishoeck09}
{Herbst}, E. \& {van Dishoeck}, E.~F. 2009, \araa, 47, 427

\bibitem[{{Jim{\'e}nez-Serra} {et~al.}(2008){Jim{\'e}nez-Serra}, {Caselli}, {Mart{\'\i}n-Pintado}, \& {Hartquist}}]{Jimenez-Serra08}
{Jim{\'e}nez-Serra}, I., {Caselli}, P., {Mart{\'\i}n-Pintado}, J., \& {Hartquist}, T.~W. 2008, \aap, 482, 549

\bibitem[{{Knacke} {et~al.}(1973){Knacke}, {Strom}, {Strom}, {Young}, \& {Kunkel}}]{Knacke73}
{Knacke}, R.~F., {Strom}, K.~M., {Strom}, S.~E., {Young}, E., \& {Kunkel}, W. 1973, \apj, 179, 847

\bibitem[{{Levreault}(1988)}]{Levreault88}
{Levreault}, R.~M. 1988, \apjs, 67, 283

\bibitem[{{Lindberg} \& {J{\o}rgensen}(2012)}]{LindbergJorgensen12}
{Lindberg}, J.~E. \& {J{\o}rgensen}, J.~K. 2012, \aap, 548, A24

\bibitem[{{Lindberg} {et~al.}(2014){Lindberg}, {J{\o}rgensen}, {Brinch}, {Haugb{\o}lle}, {Bergin}, {Harsono}, {Persson}, {Visser}, \& {Yamamoto}}]{Lindberg14}
{Lindberg}, J.~E., {J{\o}rgensen}, J.~K., {Brinch}, C., {et~al.} 2014, \aap, 566, A74

\bibitem[{{Lindberg} {et~al.}(2015){Lindberg}, {J{\o}rgensen}, {Watanabe}, {Bisschop}, {Sakai}, \& {Yamamoto}}]{Lindberg15}
{Lindberg}, J.~E., {J{\o}rgensen}, J.~K., {Watanabe}, Y., {et~al.} 2015, \aap, 584, A28

\bibitem[{{Liu} {et~al.}(2014){Liu}, {Galv{\'a}n-Madrid}, {Forbrich}, {Rodr{\'\i}guez}, {Takami}, {Costigan}, {Manara}, {Yan}, {Karr}, {Chou}, {Ho}, \& {Zhang}}]{Liu14}
{Liu}, H.~B., {Galv{\'a}n-Madrid}, R., {Forbrich}, J., {et~al.} 2014, \apj, 780, 155

\bibitem[{{Miettinen} {et~al.}(2008){Miettinen}, {Kontinen}, {Harju}, \& {Higdon}}]{Miettinen08}
{Miettinen}, O., {Kontinen}, S., {Harju}, J., \& {Higdon}, J.~L. 2008, \aap, 486, 799

\bibitem[{{Miotello} {et~al.}(2014){Miotello}, {Testi}, {Lodato}, {Ricci}, {Rosotti}, {Brooks}, {Maury}, \& {Natta}}]{Miotello14}
{Miotello}, A., {Testi}, L., {Lodato}, G., {et~al.} 2014, \aap, 567, A32

\bibitem[{{M{\"u}ller} {et~al.}(2005){M{\"u}ller}, {Schl{\"o}der}, {Stutzki}, \& {Winnewisser}}]{Muller05}
{M{\"u}ller}, H. S.~P., {Schl{\"o}der}, F., {Stutzki}, J., \& {Winnewisser}, G. 2005, Journal of Molecular Structure, 742, 215

\bibitem[{{Neuh{\"a}user} \& {Forbrich}(2008)}]{Neuhauser08}
{Neuh{\"a}user}, R. \& {Forbrich}, J. 2008, in Handbook of Star Forming Regions, Volume II, ed. B.~{Reipurth}, Vol.~5, 735

\bibitem[{{Nutter} {et~al.}(2005){Nutter}, {Ward-Thompson}, \& {Andr{\'e}}}]{Nutter05}
{Nutter}, D.~J., {Ward-Thompson}, D., \& {Andr{\'e}}, P. 2005, \mnras, 357, 975

\bibitem[{{Ohashi} {et~al.}(2023){Ohashi}, {Tobin}, {J{\o}rgensen}, {Takakuwa}, {Sheehan}, {Aikawa}, {Li}, {Looney}, {Williams}, {Aso}, {Sharma}, {Sai Insa Choi}, {Yamato}, {Lee}, {Tomida}, {Yen}, {Encalada}, {Flores}, {Gavino}, {Kido}, {Han}, {Lin}, {Narayanan}, {Phuong}, {Santamar{\'\i}a-Miranda}, {Thieme}, {van't Hoff}, {de Gregorio-Monsalvo}, {Koch}, {Kwon}, {Lai}, {Lee}, {Plunkett}, {Saigo}, {Hirano}, {Lam}, \& {Mori}}]{Ohashi23}
{Ohashi}, N., {Tobin}, J.~J., {J{\o}rgensen}, J.~K., {et~al.} 2023, \apj, 951, 8

\bibitem[{{Okoda} {et~al.}(2021){Okoda}, {Oya}, {Francis}, {Johnstone}, {Inutsuka}, {Ceccarelli}, {Codella}, {Chandler}, {Sakai}, {Aikawa}, {Alves}, {Balucani}, {Bianchi}, {Bouvier}, {Caselli}, {Caux}, {Charnley}, {Choudhury}, {De Simone}, {Dulieu}, {Dur{\'a}n}, {Evans}, {Favre}, {Fedele}, {Feng}, {Fontani}, {Hama}, {Hanawa}, {Herbst}, {Hirota}, {Imai}, {Isella}, {J{\'\i}menez-Serra}, {Kahane}, {Lefloch}, {Loinard}, {L{\'o}pez-Sepulcre}, {Maud}, {Maureira}, {Menard}, {Mercimek}, {Miotello}, {Moellenbrock}, {Mori}, {Murillo}, {Nakatani}, {Nomura}, {Oba}, {O'Donoghue}, {Ohashi}, {Ospina-Zamudio}, {Pineda}, {Podio}, {Rimola}, {Sakai}, {Segura-Cox}, {Shirley}, {Svoboda}, {Taquet}, {Testi}, {Vastel}, {Viti}, {Watanabe}, {Watanabe}, {Witzel}, {Xue}, {Zhang}, {Zhao}, \& {Yamamoto}}]{Okoda21}
{Okoda}, Y., {Oya}, Y., {Francis}, L., {et~al.} 2021, \apj, 910, 11

\bibitem[{{Ossenkopf} \& {Henning}(1994)}]{Ossenkopf94}
{Ossenkopf}, V. \& {Henning}, T. 1994, \aap, 291, 943

\bibitem[{{Perotti} {et~al.}(2023){Perotti}, {J{\o}rgensen}, {Rocha}, {Plunkett}, {de la Villarmois}, {Kristensen}, {Sewi{\l}o}, {Bjerkeli}, {Fraser}, \& {Charnley}}]{Perotti23}
{Perotti}, G., {J{\o}rgensen}, J.~K., {Rocha}, W.~R.~M., {et~al.} 2023, \aap, 678, A78

\bibitem[{Peterson {et~al.}(2011)Peterson, o~Garatti, Bourke, Forbrich, Gutermuth, Jørgensen, Allen, Patten, Dunham, Harvey, Merín, Chapman, Cieza, Huard, Knez, Prager, \& Evans}]{Peterson11}
Peterson, D.~E., o~Garatti, A.~C., Bourke, T.~L., {et~al.} 2011, ApJS, 194, 43

\bibitem[{{Podio} {et~al.}(2021){Podio}, {Tabone}, {Codella}, {Gueth}, {Maury}, {Cabrit}, {Lefloch}, {Maret}, {Belloche}, {Andr{\'e}}, {Anderl}, {Gaudel}, \& {Testi}}]{Podio21}
{Podio}, L., {Tabone}, B., {Codella}, C., {et~al.} 2021, \aap, 648, A45

\bibitem[{{Rabli} \& {Flower}(2010)}]{RabliFlower10}
{Rabli}, D. \& {Flower}, D.~R. 2010, \mnras, 406, 95

\bibitem[{{Sabatini} {et~al.}(2021){Sabatini}, {Bovino}, {Giannetti}, {Grassi}, {Brand}, {Schisano}, {Wyrowski}, {Leurini}, \& {Menten}}]{Sabatini21}
{Sabatini}, G., {Bovino}, S., {Giannetti}, A., {et~al.} 2021, \aap, 652, A71

\bibitem[{{Sabatini} {et~al.}(2023){Sabatini}, {Bovino}, \& {Redaelli}}]{Sabatini23}
{Sabatini}, G., {Bovino}, S., \& {Redaelli}, E. 2023, \apjl, 947, L18

\bibitem[{{Sabatini} {et~al.}(2022){Sabatini}, {Bovino}, {Sanhueza}, {Morii}, {Li}, {Redaelli}, {Zhang}, {Lu}, {Feng}, {Tafoya}, {Izumi}, {Sakai}, {Tatematsu}, \& {Allingham}}]{Sabatini22}
{Sabatini}, G., {Bovino}, S., {Sanhueza}, P., {et~al.} 2022, \apj, 936, 80

\bibitem[{{Sabatini} {et~al.}(2019){Sabatini}, {Giannetti}, {Bovino}, {Brand}, {Leurini}, {Schisano}, {Pillai}, \& {Menten}}]{Sabatini19}
{Sabatini}, G., {Giannetti}, A., {Bovino}, S., {et~al.} 2019, \mnras, 490, 4489

\bibitem[{Sandell {et~al.}(2021)Sandell, Reipurth, Vacca, \& Bajaj}]{Sandell21}
Sandell, G., Reipurth, B., Vacca, W.~D., \& Bajaj, N.~S. 2021, \apj, 920, 7

\bibitem[{{Sanhueza} {et~al.}(2019){Sanhueza}, {Contreras}, {Wu}, {Jackson}, {Guzm{\'a}n}, {Zhang}, {Li}, {Lu}, {Silva}, {Izumi}, {Liu}, {Miura}, {Tatematsu}, {Sakai}, {Beuther}, {Garay}, {Ohashi}, {Saito}, {Nakamura}, {Saigo}, {Veena}, {Nguyen-Luong}, \& {Tafoya}}]{Sanhueza19}
{Sanhueza}, P., {Contreras}, Y., {Wu}, B., {et~al.} 2019, \apj, 886, 102

\bibitem[{{Santos} {et~al.}(2022){Santos}, {Chuang}, {Lamberts}, {Fedoseev}, {Ioppolo}, \& {Linnartz}}]{Santos22}
{Santos}, J.~C., {Chuang}, K.-J., {Lamberts}, T., {et~al.} 2022, \apjl, 931, L33

\bibitem[{{Sch{\"o}ier} {et~al.}(2005){Sch{\"o}ier}, {van der Tak}, {van Dishoeck}, \& {Black}}]{Schoier05}
{Sch{\"o}ier}, F.~L., {van der Tak}, F.~F.~S., {van Dishoeck}, E.~F., \& {Black}, J.~H. 2005, \aap, 432, 369

\bibitem[{{Shirley} {et~al.}(2005){Shirley}, {Nordhaus}, {Grcevich}, {Evans}, {Rawlings}, \& {Tatematsu}}]{Shirley05}
{Shirley}, Y.~L., {Nordhaus}, M.~K., {Grcevich}, J.~M., {et~al.} 2005, \apj, 632, 982

\bibitem[{{Sicilia-Aguilar} {et~al.}(2013){Sicilia-Aguilar}, {Henning}, {Linz}, {Andr{\'e}}, {Stutz}, {Eiroa}, \& {White}}]{Sicilia-Aguilar13}
{Sicilia-Aguilar}, A., {Henning}, T., {Linz}, H., {et~al.} 2013, \aap, 551, A34

\bibitem[{{Sissa} {et~al.}(2019){Sissa}, {Gratton}, {Alcal{\`a}}, {Desidera}, {Messina}, {Mesa}, {D'Orazi}, \& {Rigliaco}}]{Sissa19}
{Sissa}, E., {Gratton}, R., {Alcal{\`a}}, J.~M., {et~al.} 2019, \aap, 630, A132

\bibitem[{{Takakuwa} {et~al.}(2024){Takakuwa}, {Saigo}, {Kido}, {Ohashi}, {Tobin}, {J{\o}rgensen}, {Aikawa}, {Aso}, {Gavino}, {Han}, {Koch}, {Kwon}, {Lee}, {Lee}, {Li}, {Lin}, {Looney}, {Mori}, {Sai}, {Sharma}, {Sheehan}, {Tomida}, {Williams}, {Yamato}, \& {Yen}}]{Takakuwa24}
{Takakuwa}, S., {Saigo}, K., {Kido}, M., {et~al.} 2024, arXiv e-prints, arXiv:2401.08722

\bibitem[{{Testi} {et~al.}(2014){Testi}, {Birnstiel}, {Ricci}, {Andrews}, {Blum}, {Carpenter}, {Dominik}, {Isella}, {Natta}, {Williams}, \& {Wilner}}]{Testi14}
{Testi}, L., {Birnstiel}, T., {Ricci}, L., {et~al.} 2014, in Protostars and Planets VI, ed. H.~{Beuther}, R.~S. {Klessen}, C.~P. {Dullemond}, \& T.~{Henning}, 339--361

\bibitem[{{The CASA Team} {et~al.}(2022){The CASA Team}, Bean, Bhatnagar, Castro, Meyer, Emonts, Garcia, Garwood, Golap, Villalba, Harris, Hayashi, Hoskins, Hsieh, Jagannathan, Kawasaki, Keimpema, Kettenis, Lopez, Marvil, Masters, McNichols, Mehringer, Miel, Moellenbrock, Montesino, Nakazato, Ott, Petry, Pokorny, Raba, Rau, Schiebel, Schweighart, Sekhar, Shimada, Small, Steeb, Sugimoto, Suoranta, Tsutsumi, van Bemmel, Verkouter, Wells, Xiong, Szomoru, Griffith, Glendenning, \& Kern}]{CASA_Team_22}
{The CASA Team}, Bean, B., Bhatnagar, S., {et~al.} 2022, PASP, 134, 114501

\bibitem[{{Tsukamoto} {et~al.}(2021){Tsukamoto}, {Machida}, \& {Inutsuka}}]{Tsukamoto21}
{Tsukamoto}, Y., {Machida}, M.~N., \& {Inutsuka}, S. 2021, \apj, 913, 148

\bibitem[{{van Kempen} {et~al.}(2009){van Kempen}, {van Dishoeck}, {Hogerheijde}, \& {G{\"u}sten}}]{vanKempen09}
{van Kempen}, T.~A., {van Dishoeck}, E.~F., {Hogerheijde}, M.~R., \& {G{\"u}sten}, R. 2009, \aap, 508, 259

\bibitem[{{Watanabe} \& {Kouchi}(2002)}]{WatanabeKouchi02}
{Watanabe}, N. \& {Kouchi}, A. 2002, \apjl, 571, L173

\bibitem[{{Watanabe} {et~al.}(2012){Watanabe}, {Sakai}, {Lindberg}, {J{\o}rgensen}, {Bisschop}, \& {Yamamoto}}]{Watanabe12}
{Watanabe}, Y., {Sakai}, N., {Lindberg}, J.~E., {et~al.} 2012, \apj, 745, 126

\bibitem[{{Wiesenfeld} \& {Faure}(2013)}]{Wiesenfeld13}
{Wiesenfeld}, L. \& {Faure}, A. 2013, \mnras, 432, 2573

\bibitem[{{Wirstr{\"o}m} {et~al.}(2011){Wirstr{\"o}m}, {Geppert}, {Hjalmarson}, {Persson}, {Black}, {Bergman}, {Millar}, {Hamberg}, \& {Vigren}}]{Wirstrom11}
{Wirstr{\"o}m}, E.~S., {Geppert}, W.~D., {Hjalmarson}, {\r{A}}., {et~al.} 2011, \aap, 533, A24

\bibitem[{{Wong} {et~al.}(2016){Wong}, {Hirashita}, \& {Li}}]{Wong16}
{Wong}, Y. H.~V., {Hirashita}, H., \& {Li}, Z.-Y. 2016, \pasj, 68, 67

\end{thebibliography}

\begin{appendix}
\section{Details on data reduction}\label{App:data}
\begin{figure}
   \centering
   \includegraphics[width=0.9\hsize]{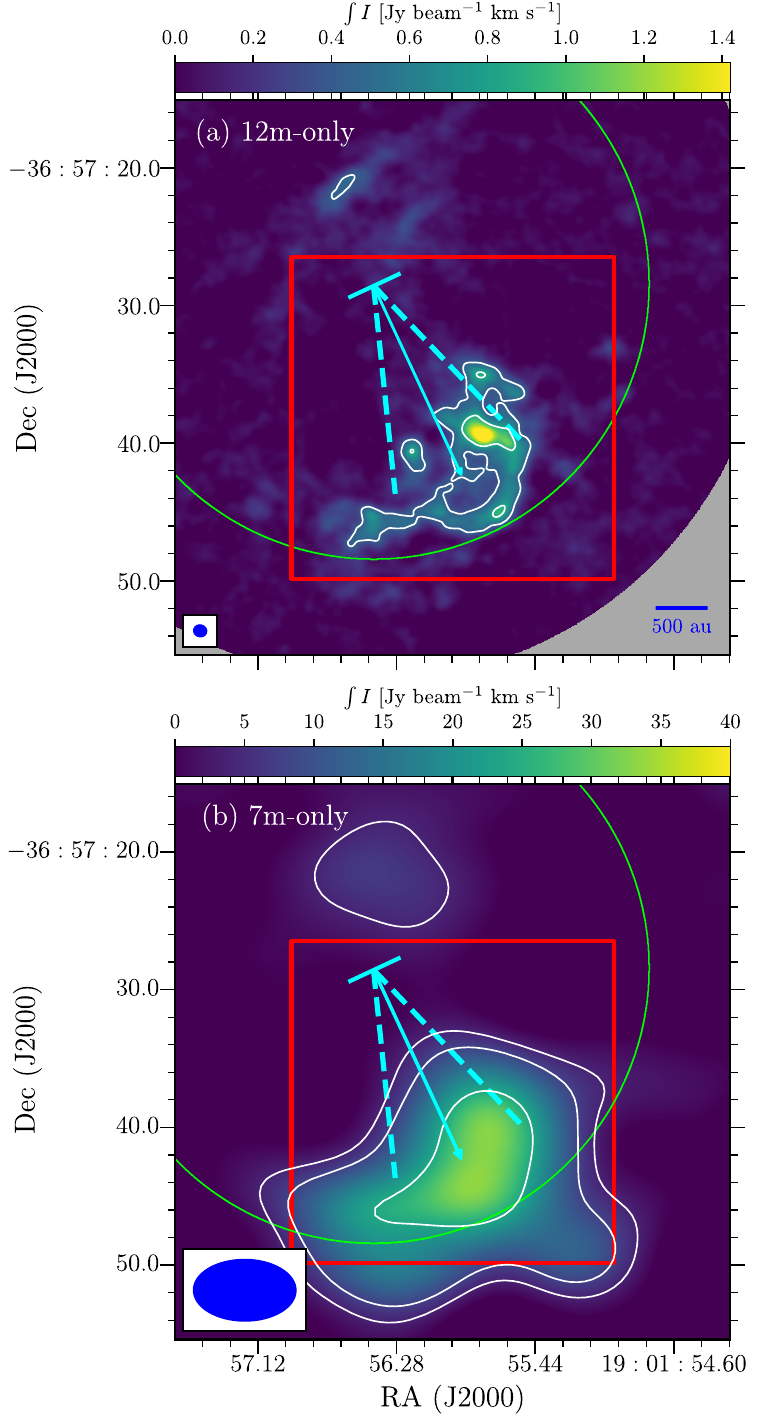}
\caption{Moment 0 of CH$_3$OH (4$_{2,3}$-3$_{1,2}$)~E as observed with (a) the 12m ALMA main array, and (b) with ACA only.
The white contours mark the [5, 10, 30]$\sigma$ emission. The green semicircle shows the ALMA Band 6 FoV of the combined data (see Figure~\ref{fig:moments_zero}). The grey background delimits the
region inside each ALMA pointing, while the red square outline the region shown in each panel in Figure~\ref{fig:moments_zero}. The ALMA beam is shown in the lower left corners.}\label{fig:12m7m}%
\end{figure}

Table~\ref{tab:obs_ALMA} summarises the details of the dataset acquired with ALMA for the CrA stellar cluster in the context of the FAUST Large Program \citep[see][]{Codella21}.  The B6 data were acquired over 12 spectral windows (SPWs) with a bandwidth~($\Delta\nu$)/frequency resolution~($\delta\nu$) of 59~MHz/122~kHz ($\sim$67-82~km~s$^{-1}$/0.17-0.20~km~s$^{-1}$), whilst 6 SPWs of 59~MHz/61~kHz ($\sim$160-190~km~s$^{-1}$/0.20~km~s$^{-1}$) were observed in B3. In all spectral settings, one additional SPW of 1.9 GHz was dedicated to the thermal dust continuum emission, i.e. 2294 (B6s1), 2404 (B6s2), and 5917~km~s$^{-1}$ (B3). For the latter, $\delta\nu$ was 0.5~MHz for B6s1 (0.72~km~s$^{-1}$) and B3 (1.54~km~s$^{-1}$), and 1.1 MHz (1.39~km~s$^{-1}$) for B6s2. Details of the observations and of the calibrators are summarised in Table~\ref{tab:obs_ALMA}. The average angular scales covered at these frequencies range from a synthesised beam, $\theta_{\rm res}$, of $\sim$0\farcs4$\times$0\farcs3 to a maximum recoverable scale, $\theta_{\rm MRS}$, of $\sim$21\arcsec~; that is, $\sim$52-2700~au at the source distance of 130 pc (\citealt{Lindberg14}). The final datacubes were obtained combining different ALMA configurations, including the 12m array and the ACA. Figure~\ref{fig:12m7m} shows an example of how the distribution of CH$_3$OH (4$_{2,3}$-3$_{1,2}$)~E, observed in B6-s1, varies with the array used.\\
\indent After the data were calibrated (Sect.~\ref{sec2:sample}), we followed a two-step deconvolution procedure for imaging: ($i$) An initial deconvolution was performed with a relatively low threshold ($\sim$4.5 times the dirty noise) and without masking the emission. A reference mask was generated applying the same threshold to the image obtained in this way. ($ii$) Then, a second deconvolution cycle was performed using the mask generated in step-$i$ as starting point for the \textsc{auto-multithresh} option in \textsc{tclean}. This two-step procedure ensures that all channels, including those with strong spatial filtering, have been properly masked.

\begin{table}[b!]
        \caption{Technical details of the FAUST observations.}\label{tab:obs_ALMA}
        \setlength{\tabcolsep}{10pt}
        \renewcommand{\arraystretch}{1}
        \renewcommand{\tabcolsep}{2pt}
        \centering
        \begin{tabular}{l|ccc}
                \toprule
                Fields & Band~3 & Band~6 & Band~6\\
                   & (B3) & (B6s1) & (B6s2)\\
                \midrule
\multirow{2}{*}{$\nu$~range~(GHz)}                        &  85.0–89.0,      & 214.0–219.0,     & 242.5–247.5, \\
                                                          &  97.0–101.0      & 229.0–234.0      & 257.2–262.5  \\
Configurations                                            &  C3-C6           & C1-C4            & C1-C4            \\
B$_{\rm min}$-B$_{\rm max}$                               &  9--1400~m       & 9--1400~m        & 15--3600~m       \\
ACA                                                       &  No              & Yes              & Yes              \\
Antennas$^{(a)}$                       &  47              & 45--48(12)       & 45--48(12)       \\
\multirow{2}{*}{$\Delta\nu$ (SPW-lines)}$^{(b)}$& 59~MHz & 59~MHz & 59~MHz \\ 
                                                          & $\sim$175~km~s$^{-1}$ & $\sim$75~km~s$^{-1}$ & $\sim$75~km~s$^{-1}$ \\ 
\multirow{2}{*}{$\delta\nu$ (SPW-lines)}$^{(b)}$& 61~kHz & 122~kHz & 122~kHz \\ 
                                                          & $\sim$0.2~km~s$^{-1}$ & $\sim$0.17~km~s$^{-1}$ & $\sim$0.17~km~s$^{-1}$ \\ 
$\langle\theta_{\rm res}\rangle$$^{(c)}$ & $0.40\arcsec\times0.27\arcsec$ & $0.42\arcsec\times0.34\arcsec$ & $0.41\arcsec\times0.30\arcsec$ \\
$\langle\theta_{\rm MRS}\rangle$$^{(c)}$ &  27$\arcsec$     & 19$\arcsec$      & 17$\arcsec$      \\
\multirow{3}{*}{Calibrators}                              &  J1924-2914,     & J1924-2914,      & J1802-3940,      \\
                                                          &  J1925-3401,     & J2056-4714,      & J1924-2914,      \\
                                                          &  J1937-3958\:    & J1957-3845\:     & J2056-4714\:     \\

                \bottomrule     
        \end{tabular}
        \tablefoot{$^{(a)}$ Parenthesis refer to the number of antennas for ACA. Ranges are given when multiple execution blocks are employed; $^{(b)}$ $\Delta\nu$ and $\delta\nu$ are the typical bandwidth and spectral resolution of SPWs dedicated to line observations; $^{(c)}$ Synthesised beam ($\theta_{\rm res} = \lambda/{\rm B_{max}}$) and maximum recoverable scale ($\theta_{\rm MRS}= 0.6\:\lambda/{\rm B_{min}}$) in the combined 12~m and ACA data sets;}
\end{table}
\section{Additional notes and results}\label{app:channels}

\begin{figure*}
  \centering
  \begin{subfigure}[b]{0.85\textwidth}
    \centering
    \includegraphics[width=0.9\linewidth]{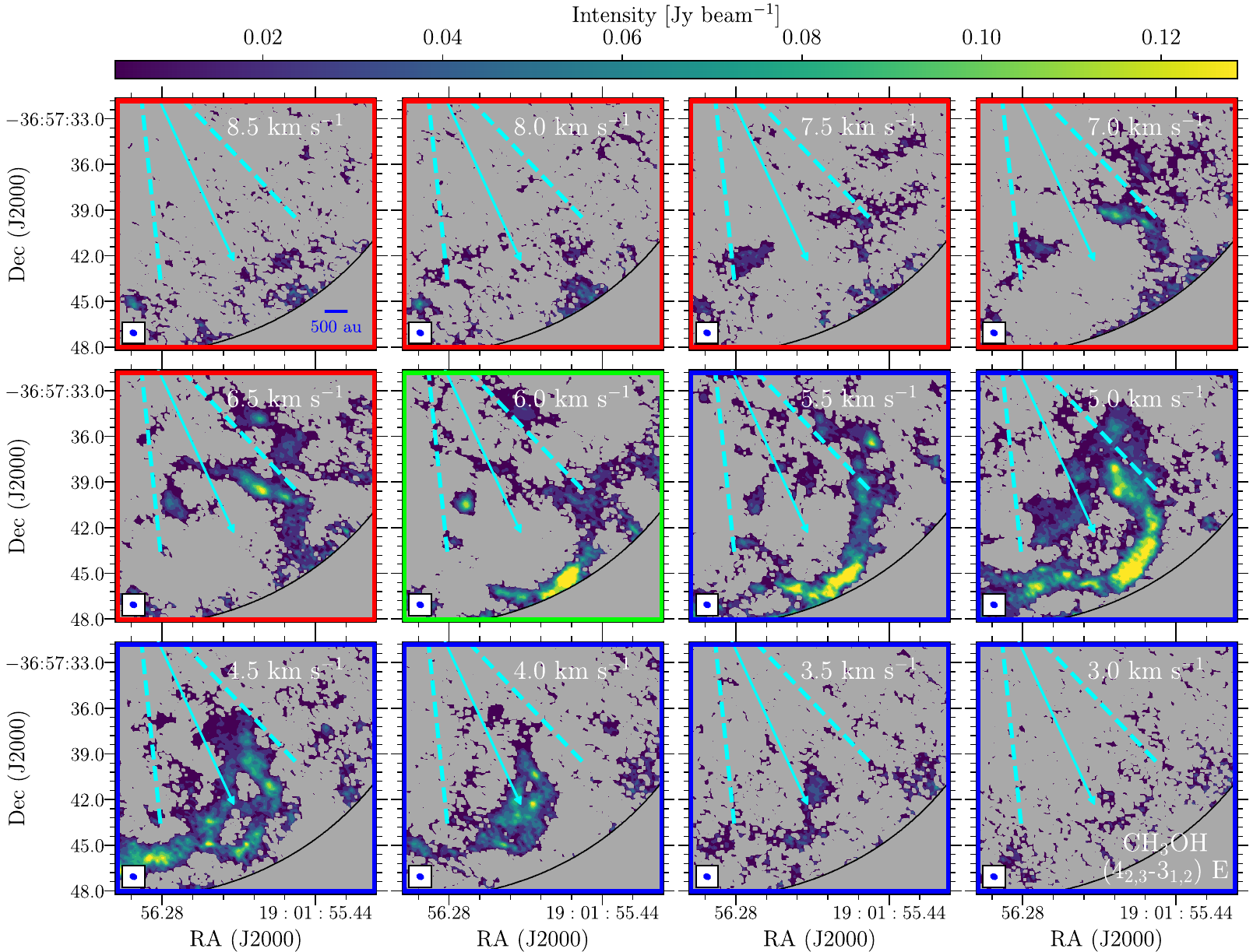}
    \caption{Channel map of CH$_3$OH (4$_{2,3}$-3$_{1,2}$) E}
  \end{subfigure}%
  \quad
  \begin{subfigure}[b]{0.85\textwidth}
    \centering
    \includegraphics[width=0.9\linewidth]{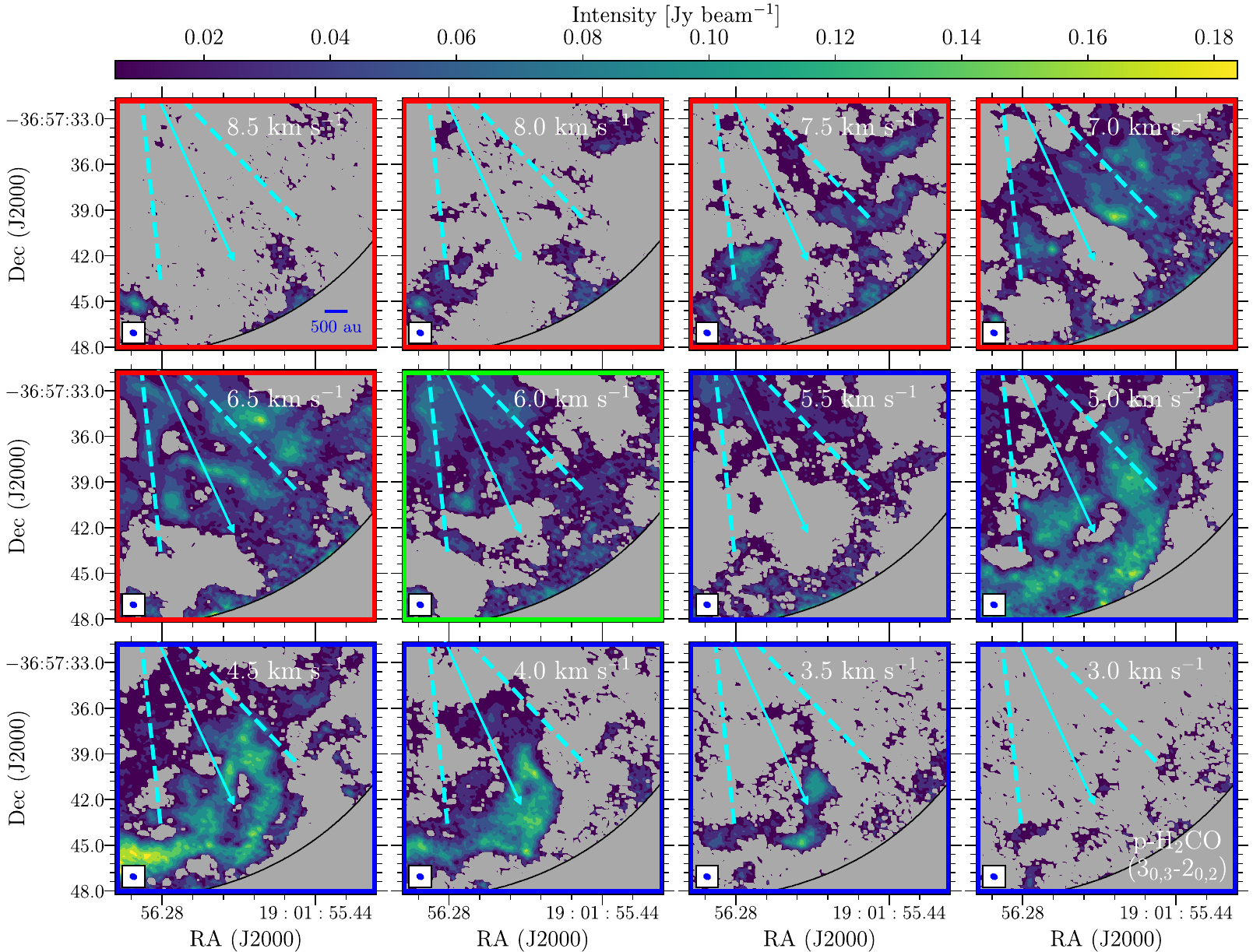}
    \caption{Channel map of p-H$_2$CO (3$_{0,3}$-2$_{0,2}$)}
  \end{subfigure}
  \caption{Velocity channel maps of (a) CH$_3$OH (4$_{2,3}$-3$_{1,2}$)~E and (b) p-H$_2$CO (3$_{0,3}$-2$_{0,2}$). The green framed panel marks the systemic velocity of IRS7B, while the blue- and red- framed panels represent the blueshifted and redshifted components of the emission, respectively. The emission was masked at 3$\sigma$, and the colour bar peaks at 75$\sigma$, where the average 1$\sigma$ noise in each channel is (a) $1.2$~mJy~beam$^{-1}$ and (b) $2$~mJy~beam$^{-1}$. The black arc shows the ALMA Band~6 FoV. Each panel reports the corresponding velocity.}\label{fig:channelmaps}
\end{figure*}

\subsection{Source properties at 1.3~mm}
We used the 2D Gaussian fit algorithm provided by CASA (\citealt{CASA_Team_22}) to analyse the 1.3 mm continuum maps (see Section~\ref{sec3.1:cont}) and extract the properties of each source identified in the \RCrA cluster.\\
\indent The fitting algorithm provided the positions of the continuum peaks of each source, the intensity of the continuum peaks, $\mathcal{F}_{\rm 1.3mm}^{\rm max}$, and the integrated flux density, $\int \mathcal{F}_{\rm 1.3mm}$~d$\Omega$, over a circular region encompassing the entire 3$\sigma$ continuum emission. A summary of these parameters is presented in Table~\ref{tab:continuum}.

\begin{table}
        \caption{Source properties derived at 1.3~mm.}\label{tab:continuum}
        \setlength{\tabcolsep}{10pt}
        \renewcommand{\arraystretch}{1}
        \renewcommand{\tabcolsep}{2pt}
        \centering
        \begin{tabular}{l|cccc}
                \toprule
            Source-ID   & $\alpha_{\rm ICRS}$ & $\delta_{\rm ICRS}$ & $\mathcal{F}_{\rm 1.3mm}^{\rm max}$ & $\int \mathcal{F}_{\rm 1.3mm}$~d$\Omega$\\
            & \footnotesize{hh:mm:ss.ss} & \footnotesize{dd:mm:ss.ss} & \footnotesize{mJy/beam} & \footnotesize{mJy}\\
            \midrule
            IRS7B   & 19:01:56.42 & -36:57:28.58 & 190 &   304 \\
            SMM 1C  & 19:01:55.30 & -36:57:17.23 &  97 &   269 \\
            IRS7A   & 19:01:55.33 & -36:57:22.60 &  17 &    20 \\
            CXO 34  & 19:01:55.79 & -36:57:28.23 &  16 &    16 \\
            FAUST-5 & 19:01:56.63 & -36:57:40.49 &   4 &     4 \\
                \bottomrule     
        \end{tabular}
\end{table}

\begin{figure}
   \centering
   \includegraphics[width=0.95\hsize]{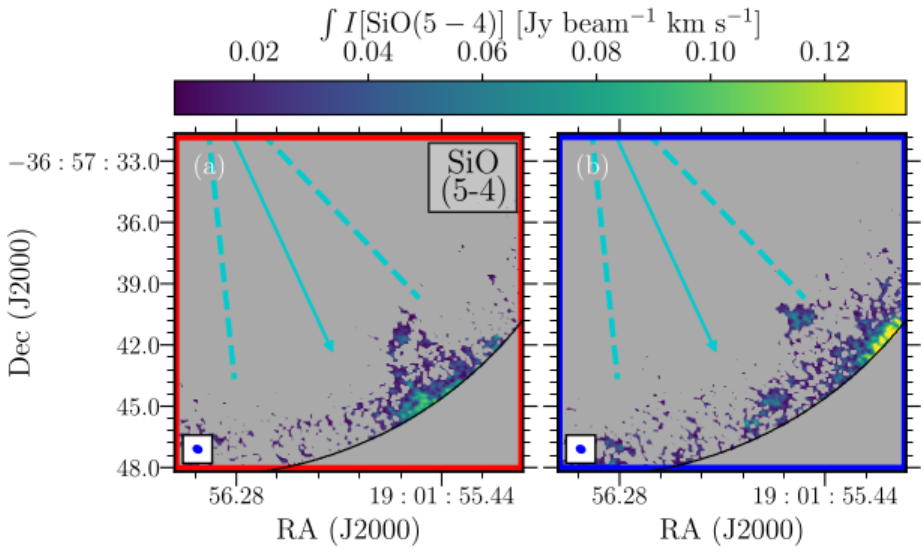}
\caption{Integrated intensity maps of the blueshifted and redshifted components of SiO emission (panels a and b, respectively). The blueshifted and redshifted components were derived by integrating the emission at $ > 3\sigma$ in the velocity ranges of $\sim$[6.7, +12]~km~s$^{-1}$ and $\sim$[0, +6.2]~km~s$^{-1}$, respectively. The black arc shows the ALMA Band~6 FoV.}\label{fig:channel_SOSiO}%
\end{figure}

\begin{figure*}[b!]
   \centering
   \includegraphics[width=0.80\hsize]{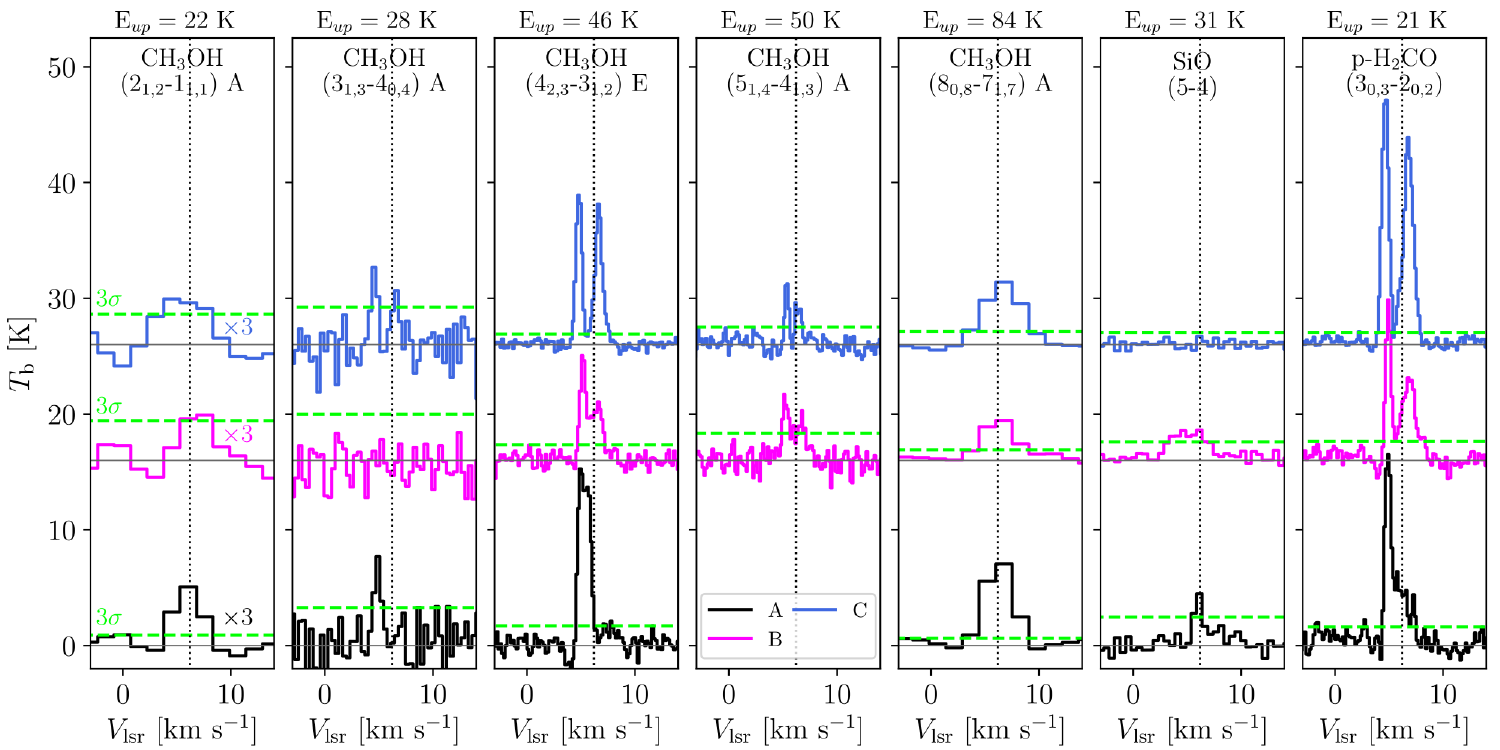}
\caption{Spectral overview of the tracers listed in Table~\ref{tab:line_properties} extracted at the three positions labeled A, B, and C in Fig.~\ref{fig:moments_zero}a. The extraction regions have an equivalent area of nine beams. Vertical lines report the \VirsB (+6 km~s$^{-1}$; \citealt{Lindberg14}, and \citealt{Ohashi23}), grey lines mark the zero-level of each spectrum, while green dashed lines show the 3$\sigma$ level.}\label{fig:spectra}
\end{figure*}

\subsection{Channel maps, velocity structures, and spectra}
In this subsection we report the channel maps of CH$_3$OH (4$_{2,3}$-3$_{1,2}$) and p-H$_2$CO (3$_{0,3}$-2$_{0,2}$) and the integrated intensity maps of the blueshifted and redshifted components of SiO~(5-4) emission, respectively, in Figures~\ref{fig:channelmaps}(a,b), and \ref{fig:channel_SOSiO}. The velocity structures defined by the three molecular tracers are masked at the 3$\sigma$ level. The green framed panels in Figure~\ref{fig:channelmaps} indicate the systemic velocity of IRS7B, while the blue- and red-framed panels represent the blueshifted and redshifted components of the emission, respectively. The same colour code was used in Figure~\ref{fig:channel_SOSiO}.\\
\indent Figures~\ref{fig:channelmaps}(a,b) show one velocity channel in steps of three original spectral resolution elements, while in Figure~\ref{fig:channel_SOSiO} we show the integrated blueshifted and redshifted components of SiO~(5-4) in the velocity ranges of $\sim$[+6.7, +12]~km~s$^{-1}$ and $\sim$[0, +6.2]~km~s$^{-1}$, respectively. Cyan lines and arrows follow Figure~\ref{fig:cont_RCrA}, and trace the supposed outflow cavity walls detected at 1.3~mm with ALMA (Section~\ref{sec4:discussion_conclusions}). Notably, the region shown in Figure~\ref{fig:channel_SOSiO} is the only part of the ALMA FoV where extended integrated SiO emission is observed.\\
\indent Figure~\ref{fig:spectra} shows the spectra of all the observed lines extracted at three positions along the arc structure, centred at the peak positions of CH$_3$OH (4$_{2,3}$-3$_{1,2}$), SiO (5-4) and p-H$_2$CO (3$_{0,3}$-2$_{0,2}$), labelled A, B and C, respectively. {The targeted lines are detected with a signal-to-noise ratio of $>$3$\sigma$. The exceptions are the CH$_3$OH-A~(3$_{1,3}$-4$_{0,4}$) at position B and the SiO (5-4) at position C, where we derived 3$\sigma$ upper limits (Table~\ref{tab:line_properties}). All the CH$_3$OH and H$_2$CO lines observed with a spectral resolution of $\leq$ 0.5 km~s$^{-1}$ show double-peaked profiles, with the redshifted component associated with extended envelope emission, and the blueshifted one due to the arc (see Appendix~\ref{app:channels}). We used the Python Spectroscopic Toolkit (PySpecKit; \citealt{Ginsburg22}) to fit the spectra assuming two Gaussian components. In regions B and C, the blueshifted lines have peak velocities of $\sim$+5~km~s$^{-1}$ and a narrow FWHM of $\sim$1~km~s$^{-1}$. The red components are centred at +6.6~km~s$^{-1}$, and have FWHMs of between 1 and 1.2~km~s$^{-1}$. The low-frequency methanol lines (\Eup~= 22~K, and 84~K) observed with a spectral resolution of 1.54 km~s$^{-1}$ show single Gaussian-like profiles. The lines peak between +5.7 and +6.3~km~s$^{-1}$, close to \VirsB. The detection of a single Gaussian line in the low-resolution data is due to the coarse spectral resolution. On the other hand, the line width of SiO at position B is $\sim 3.5$~km~s$^{-1}$.}\\
\indent {The integrated line intensities of the multiple CH$_3$OH transitions were used to construct rotational diagrams and quantify the column density and rotational temperature of CH$_3$OH along the molecular arc (see Sect.~\ref{sec3.3:RDiagram}).}

\end{appendix}
\end{document}